\documentclass[]{elsarticle}
\newcommand{\isfinal}{false}

\usepackage{array}
\usepackage{fixltx2e}
\usepackage{color}
\usepackage{psfrag}
\usepackage{epsfig}
\usepackage{tabularx}
\usepackage{amssymb}
\usepackage{ifthen}

\def\exeq{\mathrel{\stackrel{!}{=}}}

\def\PSNR{\mathrm{ PSNR}}

\def\punit{\, \mathrm}

\def\ve#1{{\mathchoice{\mbox{\boldmath$\displaystyle #1$}}%
		      {\mbox{\boldmath$\textstyle #1$}}%
		      {\mbox{\boldmath$\scriptstyle #1$}}%
		      {\mbox{\boldmath$\scriptscriptstyle #1$}}}} 

\def\diag{\mathrm{ diag}}

\def\diag{\mathrm{ diag}}

\def\trans{\mathsf{T}}

\newcommand{\argmin}{\mathop{\mathrm{argmin}}}
\newcommand{\argmax}{\mathop{\mathrm{argmax}}}

\newcolumntype{C}[1]{>{\centering\arraybackslash}p{#1}}

\newcolumntype{C}[1]{>{\centering\arraybackslash}p{#1}}

\journal{Journal of Visual Communication and Image Representation }
\begin{document}


\begin{frontmatter}
\title{Motion Compensated Three-Dimensional Frequency Selective Extrapolation for Improved Error Concealment in Video Communication \tnoteref{t1}}
\tnotetext[t1]{The work of J.\ Seiler and A.\ Kaup was partly supported by the German Research Foundation Deutsche Forschungsgemeinschaft (DFG) under contract number KA~926/1-2.}
\author{J{\"u}rgen~Seiler\corref{cor1}}
\ead{seiler@lnt.de}
\author{Andr{\'e}~Kaup}
\ead{kaup@lnt.de}
\address{Chair of Multimedia Communications and Signal Processing,\\University of Erlangen-Nuremberg,\\ Cauerstr. 7, 91058 Erlangen, Germany\vspace{0.75cm}}
\cortext[cor1]{Corresponding author, \mbox{phone: +49 9131 85 27103}, \mbox{fax: +49 9131 85 28849}}

\begin{abstract}
During transmission of video data over error-prone channels the risk of getting severe image distortions due to transmission errors is ubiquitous. To deal with image distortions at decoder side, error concealment is applied. This article presents Motion Compensated Three-Dimensional Frequency Selective Extrapolation, a novel spatio-temporal error concealment algorithm. The algorithm uses fractional-pel motion estimation and compensation as initial step, being followed by the generation of a model of the distorted signal. The model generation is conducted by an enhanced version of Three-Dimensional Frequency Selective Extrapolation, an existing error concealment algorithm. Compared to this existent algorithm, the proposed one yields an improvement in concealment quality of up to 1.64 dB PSNR. Altogether, the incorporation of motion compensation and the improved model generation extends the already high extrapolation quality of the underlying Frequency Selective Extrapolation, resulting in a gain of more than 3 dB compared to other well-known error concealment algorithms.\vspace{0.5cm} 
\end{abstract}
\end{frontmatter}

\vspace{1cm}
\section{Introduction}
\label{sec:introduction}

In the past years the computing power of multimedia devices has been steadily increased. In combination with the growing transport capacity for mobile transmission and transmission over the internet, the ability to record and play digital video data has become more and more popular. But by transporting digital video data over wireless channels or the internet, the risk of occurring transmission errors is ubiquitous. On wireless channels the errors may be caused e.\ g.\ by fading or the loss of synchronization. By using the packetized data transport of the internet, the main causes for transmission errors are delayed or lost data packets. No matter what actually caused the transmission error, the decoded video sequence becomes heavily distorted. Due to the transmission errors, parts of one or several video frames cannot be displayed, or only with severe distortion. The errors become even worse if a hybrid video codec as e.\ g.\ the H.264/AVC \cite{ISO/IEC2003} is used, since the errors propagate to following frames due to the predictive structure of such a video codec. A hybrid video codec uses already transmitted parts of the sequence for predicting the block being coded. Thus, the decoder can predict the block in the same manner as the encoder. But in the case of a distortion, the decoder has to form the prediction based on the distorted signal parts and therewith propagates the errors to future frames.

In order to cope with this, as shown in \cite{Stockhammer2005} and \cite{Wang1998}, modern hybrid video codecs use two strategies. On the one hand side, this is error resilience which aims at protecting the coded video data bitstream against transmission errors. Unfortunately, the protection is only possible at the expense of an increased data rate and therewith a decreased coding efficiency. On the other hand side, the second strategy is error concealment which takes effect in the case that a transmission error occurs and error resilience is not used or fails to correct the error. Error concealment aims at concealing the distortion caused by the transmission errors at the decoder side by extrapolating the signal from correctly received regions into the areas where the signal is distorted. So, the goal is to reconstruct the original signal for generating a pleasant image for presentation despite the error and for reducing error propagation. Since error concealment is not part of standardization, one can freely decide which concealment algorithm to use at the decoder, depending on its computational demands and memory requirements.

Within the scope of this article we focus on error concealment and present a powerful spatio-temporal error concealment algorithm, the Motion Compensated Three-Dimensional Frequency Selective Extrapolation. This algorithm uses motion estimation and compensation as initial part of the extrapolation process which is then followed by a model generation step. 

In the subsequent sections, first an overview of different error concealment algorithms is given. Afterwards, we outline the novel algorithm in detail, show how the model generation part can be significantly enhanced and especially point out how the novel algorithm can use motion estimation and compensation with fractional-pel accuracy. Finally, we prove the abilities of the algorithm by a comprehensive analysis of the concealment quality for coded video data. We also show how, in addition to the novel algorithm, other common temporal error concealment algorithms can be improved by using fractional-pel motion estimation and compensation.


\section{Previous Work on Spatial, Temporal and Spatio-Temporal Error Concealment}
\label{sec:previous_work}

The first group of error concealment algorithms to mention is the group that only uses spatial information from the currently considered frame for concealing the distorted areas. One member of this group is the Maximally Smooth Recovery algorithm proposed by Wang et.\ al.\ \cite{Wang1993}. This algorithm aims at extra\-polating distorted areas in such a way that a smooth transition to the correctly received neighboring samples results. Another algorithm, introduced by Sun and Kwok \cite{Sun1995}, makes use of projection onto convex sets for concealing the distortion. There, the block to be reconstructed is iteratively manipulated in the spatial and in the frequency domain in order to form an extrapolated signal that fulfills some spatial as well as some frequency constraints. Alkachouh and Bellanger \cite{Alkachouh2000} proposed another spatial error concealment algorithm working in the DCT domain. This algorithm recovers the lowpass DCT coefficients of the distorted block by means of the correctly received adjacent samples. Another spatial error concealment algorithm is the sequential error concealment proposed by Li and Orchard \cite{Li2002}. There, the distorted block iteratively is reconstructed and superimposed from different directions. A different approach is the Two-Dimensional Frequency Selective Extrapolation from \cite{Kaup2005} which iteratively generates a model of the distorted signal as a weighted superposition of two-dimensional basis functions. Besides this small selection of spatial error concealment algorithms many more exist. Some additional algorithms can be found in \cite{Tsekeridou2000} and \cite{Kung2006}.

According to \cite{Tsekeridou2000, Kung2006, Bopardikar2005}, due to the usually very high temporal correlations within one video sequence, using temporal error concealment algorithms in general is superior to using spatial ones. The group of temporal error concealment algorithms makes use of signal parts in frames that have been transmitted and decoded prior to the distorted frames. All algorithms of this group have in common that they seek an area in a previous frame to replace the distorted one in the current frame. Several of these algorithms are introduced in \cite{Tsekeridou2000, Kung2006, Bopardikar2005}. Out of this group we especially focus on the Extended Boundary Matching Algorithm (EBMA) from Lam et.\ al.\ \cite{Lam1993} and the Decoder Motion Vector Estimation (DMVE) proposed by Zhang et.\ al.\ \cite{Zhang2000a} since these two algorithms are widely used and will later serve for comparison. Both algorithms aim at determining the motion of the sequence in order to locate a shifted area in a previous frame which then is used for replacing the distorted area. For this, EBMA selects the area that minimizes the boundary error between the correctly received adjacent samples in the distorted current frame and the margin of the candidate area from the previous frame. If DMVE is used, an area of correctly received samples around the distortion is selected. Then, this area is compared to shifted versions of itself in the previous frame. The estimated motion corresponds to the area minimizing the error. After this, the distorted area is replaced by the one from the previous frame shifted according to the estimated motion. In the case of bi-temporal prediction where previous as well as future frames can be available, the best matching area can also be searched in a future frame. Both algorithms can use a full search as well as a restricted search for determining the candidates. Thereby, the restricted search only uses a subset of candidate motion vectors in order to select the best one among them. This subset can be obtained e.\ g.\ by regarding the motion vectors of the correctly received neighboring blocks.

The third group of error concealment algorithms, the spatio-temporal error concealment algorithms, exploits spatial as well as temporal information and therewith is superior to spatial and temporal error concealment according to \cite{Tsekeridou2000}. The first algorithm out of this group to mention is the Spatio-Temporal Fading Scheme from Friebe and Kaup \cite{Friebe2006a} and its extension to B-Frames \cite{Friebe2006}. This algorithm first performs a temporal and a spatial extrapolation and then conducts a weighted superposition of them. For this, the weights for the spatial and the temporal extrapolation are determined by using the boundary error. This algorithm is improved by Hwang et.\ al., leading to the Improved Fading Scheme (IFS) \cite{Hwang2008}. There, a superposition of DMVE and EBMA is used for temporal extrapolation and a superposition of stripe patch repetition and bilinear interpolation for spatial extrapolation. The Spatio-Temporal Boundary Matching Algorithm \cite{Chen2006} from Chen et.\ al.\ follows a different approach. This algorithm exploits temporal and spatial smoothness of a signal to recover a motion vector and therewith is an extension of the original EBMA. An improvement to the original DMVE is achieved by Multi-Hypothesis Error Concealment \mbox{(MH-EC) \cite{Song2007}} from Song et.\ al.. This algorithm determines several candidate blocks that can be used for concealing the distortion and superimposes them. With that, the idea of Multihypothesis Motion Estimation from Flierl and Girod \cite{Flierl2001} is applied to error concealment. The Spatio-Temporal Error Concealment with Optimized Mode Selection from Belfiore et. al.\ \cite{Belfiore2003} is another algorithm out of this group and is able to effectively switch between spatial and temporal extrapolation and a combination of them. A similar approach is proposed by Hadar et.\ al., regarding their Hybrid Error Concealment \cite{Hadar2005}, where the switch between the different initial signal extrapolations is performed differently. Another spatio-temporal error concealment algorithm is the Boundary Matching Algorithm and Mesh-Based Warping from Atzori et.\ al.\ \cite{Atzori2001}. There, first a preliminary temporal extrapolation is obtained by using the Boundary Matching Algorithm, which is spatially warped to fit the correctly received neighborhood afterwards. A completely different approach is the spatio-temporal concealment by Three-Dimensional Frequency Selective Extrapolation (\mbox{3D-FSE}) proposed by Meisinger and Kaup in \cite{Meisinger2007} which is the extension of the algorithm from \cite{Kaup2005} into three dimensions. This algorithm aims at generating a joint three-dimensional model of the signal suffering from distortion. 

Unfortunately, 3D-FSE is only able to compensate small motion effectively. In the case of large motion, the model generation is performed only suboptimally and the generated model may not fit the original signal well. In order to cope with this shortcoming, we propose the subsequently described Motion Compensated Three-Dimensional Frequency Selective Extrapolation. The algorithm is an enhancement to 3D-FSE and uses an explicit estimation of the motion in a sequence and takes this into account for the model generation.


\section{Motion Compensated Three-Dimensional Frequency Selective \\Extrapolation}
\label{sec:MC-FSE}

Before actually presenting Motion Compensated Three-Dimensional Frequency Selective Extrapolation (MC-FSE) in detail, we will briefly review the existing 3D-FSE from \cite{Meisinger2007} in the next subsection. So, we can point out its shortcoming in scenes with motion and therewith we can show the necessity of MC-FSE. In addition to that, we will also introduce an effective modification to the original model generation algorithm which is also a part of the later outlined MC-FSE.

\subsection{Three-Dimensional Frequency Selective Extrapolation}
\label{ssec:fse}

\begin{figure*}
 	\centering
\ifthenelse{\equal{\isfinal}{true}}
	{
	\psfrag{x}[c][c][0.9]{$x$}
	\psfrag{y}[c][c][0.9]{$y$}
	\psfrag{t}[c][c][0.9]{$t$}
	\psfrag{m}[c][c][0.9]{$m$}
	\psfrag{n}[c][c][0.9]{$n$}
	\psfrag{p}[c][c][0.9]{$p$}
	\psfrag{x0}[c][c][0.7]{$x_0$}
	\psfrag{y0}[c][c][0.7]{$y_0$}
	\psfrag{ttau}[c][c][0.7]{$t=\tau$}
	\psfrag{ttaum1}[c][c][0.7]{$t=\tau-1$}
	\psfrag{ttaum2}[c][c][0.7]{$t=\tau-2$}
	\psfrag{ttaum3}[c][c][0.7]{$t=\tau-3$}
	\psfrag{ttaup1}[c][c][0.7]{$t=\tau+1$}
	\psfrag{Loss area}[l][l][0.9]{Loss area $\mathcal{B}$}
	\psfrag{Support volume}[l][l][0.9]{Support volume $\mathcal{A}$}
	\psfrag{Video sequence}[l][l][0.9]{Video sequence}
	\psfrag{Extrapolation volume}[l][l][0.9]{Extrapolation volume}
	\includegraphics[width=0.83\textwidth]{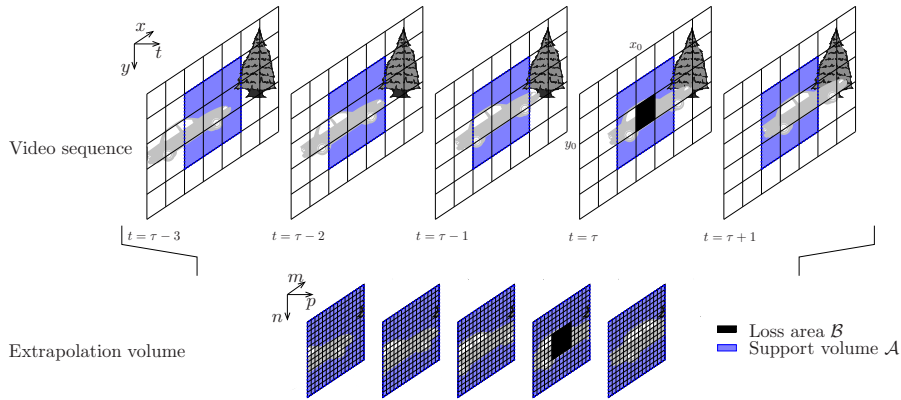}
	}{
	\psfrag{x}[c][c][0.7]{$x$}
	\psfrag{y}[c][c][0.7]{$y$}
	\psfrag{t}[c][c][0.7]{$t$}
	\psfrag{m}[c][c][0.7]{$m$}
	\psfrag{n}[c][c][0.7]{$n$}
	\psfrag{p}[c][c][0.7]{$p$}
	\psfrag{x0}[c][c][0.5]{$x_0$}
	\psfrag{y0}[c][c][0.5]{$y_0$}
	\psfrag{ttau}[c][c][0.5]{$t=\tau$}
	\psfrag{ttaum1}[c][c][0.5]{$t=\tau-1$}
	\psfrag{ttaum2}[c][c][0.5]{$t=\tau-2$}
	\psfrag{ttaum3}[c][c][0.5]{$t=\tau-3$}
	\psfrag{ttaup1}[c][c][0.5]{$t=\tau+1$}
	\psfrag{Loss area}[l][l][0.7]{Loss area $\mathcal{B}$}
	\psfrag{Support volume}[l][l][0.7]{Support volume $\mathcal{A}$}
	\psfrag{Video sequence}[l][l][0.7]{Video sequence}
	\psfrag{Extrapolation volume}[l][l][0.7]{Extrapolation volume}
	\includegraphics[width=0.95\textwidth]{graphics/3dfse-extrapolation_area_2.eps}
	}
	
 	\caption{Top: distorted video sequence $v\left[x,y,t\right]$ with isolated block loss in frame $t=\tau$ at $\left(x_0,y_0\right)$. Bottom: extrapolation volume $\mathcal{L}$ used for 3D-FSE \cite{Meisinger2007}.} 	
 	\label{fig:3dfse_extrapolation_area}
\end{figure*}

For formulating 3D-FSE, and later the MC-FSE algorithm, the following scenario is regarded: we consider the video sequence $v\left[x,y,t\right]$ which is shown in Figures \ref{fig:3dfse_extrapolation_area} and \ref{fig:mcfse_extrapolation_area} top and is depicted by spatial coordinates $x$ and $y$ and temporal coordinate $t$. The spatial size of the sequence is $X\times Y$. Without loss of generality, an isolated block loss of size $L_\mathrm{x}\times L_\mathrm{y}$ samples occurs in the frame $t=\tau$ at position $\left(x_0, y_0\right)$. The concealment of an isolated block loss is considered only for a comprehensible description. In general, the areas to conceal can be arbitrarily shaped. In such cases, as described in \cite{Kaup2005}, the area to conceal is divided into blocks and the individual blocks are concealed successively. In the course of this procedure, already concealed block are used for supporting the concealment of the next ones. 

For concealing the loss, 3D-FSE utilizes correctly received samples from the current and from previous frames and, if available, future frames. The number of involved previous frames is depicted by $N_\mathrm{p}$, of future frames by $N_\mathrm{f}$. As mentioned before, 3D-FSE aims at generating a model of the signal suffering from the distortion. For this, a data volume, the so called extrapolation volume $\mathcal{L}$, centered by the loss, is cut out of the video sequence. This volume consists of the loss of size $L_\mathrm{x}\times L_\mathrm{y}$ samples, surrounded by a border of $B$ samples width from the frame $t=\tau$. In addition to this, from the involved previous and future frames the samples at the same spatial positions are taken. The extrapolation volume $\mathcal{L}$ is depicted by coordinates $m$, $n$ and $p$ and is of size $M\times N\times P$ with
\begin{eqnarray}
 M &=& L_\mathrm{x} + 2B\\
 N &=& L_\mathrm{y} + 2B\\
 P &=& N_\mathrm{p}+N_\mathrm{f}+1.
\end{eqnarray} 
The signal in volume $\mathcal{L}$ is depicted by $f\left[m,n,p\right]$ and emanates from $v\left[x,y,t\right]$ according to
\begin{equation}
	f\left[m,n,p\right] = v\left[x_0-B+m, y_0-B+n, \tau-N_\mathrm{p}+p\right].
\end{equation} 
Further, volume $\mathcal{L}$ can be split into two parts, region $\mathcal{B}$ containing the lost samples and support volume $\mathcal{A}$, subsuming all correctly received samples used for model generation. The relation between video sequence $v\left[x,y,t\right]$ and signal $f\left[m,n,p\right]$ in volume $\mathcal{L}$ can also be seen in Fig. \ref{fig:3dfse_extrapolation_area}. 

Then, a model of the signal is generated in this volume. This model is obtained by iteratively approximating the correctly received signal parts in $\mathcal{A}$, but as the model is defined over the complete volume $\mathcal{L}$, an estimate of the lost block $\mathcal{B}$ can be achieved. Subsequently, the algorithm is outlined as sparse approximation problem. Sparsity-based algorithms are suited well for solving underdetermined problems, to which the extrapolation problem belongs. As shown in \cite{Olshausen1997}, these algorithms are able to cover important signal characteristics even if the underlying problem is underdetermined. The Frequency Selective Extrapolation belongs to the group of greedy sparse approximation algorithms as described in \cite{Tropp2004} and \cite{Temlyakov2000}. Another approximation algorithm out of this group is Matching Pursuit, originally proposed by Mallat and Zhang \cite{Mallat1993}. These iterative greedy approximation algorithms can solve sparse problems efficiently and are much more suited for the model generation than a direct solution of the sparse problem using the $L_0$-norm since this is a NP-hard problem as shown in \cite{Natarajan1995}. 

3D-FSE now aims at generating the parametric model
\begin{equation}
 g\left[m,n,p\right] = \sum_{\forall k\in \mathcal{K}} \hat{c}_k \varphi_k\left[m,n,p\right]
\end{equation}
over the complete extrapolation volume $\mathcal{L}$. The model is a weighted superposition of the mutually orthogonal basis functions $\varphi_k\left[m,n,p\right]$, and $\hat{c}_k$ denotes the individual weights. The set $\mathcal{K}$ contains the indices of all basis functions used for model generation. The model generation algorithm on the one hand side has to determine which basis functions to use for the model and on the other hand side has to calculate the corresponding weights. These two tasks are carried out in an iterative manner and, in doing so, one basis function is selected and the corresponding weight is calculated in every iteration.

For generating the signal model by 3D-FSE, the three-dimensional signals in extrapolation volume $\mathcal{L}$ are considered as vectors with the reversible mapping
\begin{equation}
	\left[\ve{x}\right]_{ m+M\cdot n+M\cdot N\cdot p} = x\left[ m,n,p\right] 
\end{equation}
\begin{eqnarray*}
\mbox{with } && m=0,\ldots ,M-1 \\ &&n=0, \ldots, N-1  \\ && p=0, \ldots, P-1.
\end{eqnarray*}
This mapping produces the vector $\ve{x}$ for any arbitrary signal $x\left[m,n,p\right]$. Expression $\left[\ve{x}\right]_i$ is used to indicate the $i$-th element of the vector $\ve{x}$. According to this, signal $f\left[m,n,p\right]$ is represented by $\ve{f}$, model $g\left[m,n,p\right]$ by $\ve{g}$ and the basis functions $\varphi_k\left[m,n,p\right]$ by basis vectors $\ve{\varphi}_k$. In addition to this, the basis vectors are subsumed in the matrix
\begin{equation}
 	\ve{\Phi} = \left(\ve{\varphi}_0, \ve{\varphi}_1, \ldots, \ve{\varphi}_{\left|\mathcal{L}\right|-1}\right).
\end{equation}
Thus, the model vector $\ve{g}$ results from
\begin{equation}
	\ve{g} = \ve{\Phi} \hat{\ve{c}}
\end{equation}
with the coefficient vector $\hat{\ve{c}}$ to be determined in the following. The coefficient vector $\hat{\ve{c}}$ is built up iteratively in such a way that the approximation error between the model and the undistorted signal parts is reduced in every iteration. For this, the residual vector $\ve{r}^{\left(0\right)}$ before the first iteration is set to 
\begin{equation}
\ve{r}^{\left(0\right)} = \ve{f}.
\end{equation}
and the coefficient vector $\hat{\ve{c}}$ is set to $\ve{0}$.

At the beginning of each iteration, here the $\nu$-th one is regarded, a weighted projection of the last iteration's approximation residual $\ve{r}^{\left(\nu-1\right)}$ onto all possible basis vectors is performed. This yields the projection vector $\ve{p}^{\left(\nu\right)}$ which can be calculated as follows:
\begin{equation}
	\ve{p}^{\left(\nu\right)} = \left(\ve{r}^{\left(\nu-1\right)}\right)^\trans \ve{W} \ve{\Phi} \left(\diag\left(\ve{\Phi}^\trans \ve{W} \ve{\Phi}\right)\right)^{-1}
\end{equation}
The weighting matrix $\ve{W} = \diag\left(\ve{w}\right)$ is used to mask the lost area and to control the influence each sample has on the model generation depending on its position. $\ve{w}$ and therewith $\ve{W}$ emanate from the three-dimensional weighting function:
\ifthenelse{\equal{\isfinal}{true}}{
\begin{equation}
	w\left[m,n,p\right] = \left\{ \begin{array}{cl} \hat{\rho}^{\sqrt{\left(m-\frac{M-1}{2}\right)^2+\left(n-\frac{N-1}{2}\right)^2+\left(p-\frac{P-1}{2}\right)^2}} & \\& \hspace{-2cm} \forall \left(m,n,p\right) \in \mathcal{A} \\ 0 & \hspace{-2cm} \forall \left(m,n,p\right) \in \mathcal{B} \end{array}\right. 
	\label{eq:mc-fse_weighting_matrix}
\end{equation}
}{
\begin{equation}
	w\left[m,n,p\right] = \left\{ \begin{array}{cl} \hat{\rho}^{\sqrt{\left(m-\frac{M-1}{2}\right)^2+\left(n-\frac{N-1}{2}\right)^2+\left(p-\frac{P-1}{2}\right)^2}} &  \forall \left(m,n,p\right) \in \mathcal{A} \\ 0 & \forall \left(m,n,p\right) \in \mathcal{B} \end{array}\right. 
	\label{eq:mc-fse_weighting_matrix}
\end{equation}
}
According to \cite{Meisinger2007}, the samples in the support volume are weighted by an iso\-tropic model with decay factor $\hat{\rho}$. With that, samples get lower weight and consequently less influence on the model generation with increasing distance to the lost block. In the case that consecutive block losses are considered, the already concealed blocks are used for concealment as well. For this purpose the region corresponding to the already concealed blocks is further weighted by the reliability coefficient $\delta$ which is smaller than $1$.

After having determined projection vector $\ve{p}^{\left(\nu\right)}$, one basis vector has to be selected to be added to the model in this iteration. For this, the vector is selected that minimizes the weighted distance between the residual $\ve{r}^{\left(\nu-1\right)}$ and the weighted projection onto the corresponding basis vector. So the index $u$ of the basis vector to be added can be determined by
\ifthenelse{\equal{\isfinal}{true}}
{
\begin{eqnarray}
 	\nonumber u \hspace{-1mm}&=&\hspace{-1mm} \argmin_k   \bigg(\left( \ve{r}^{\left(\nu-1\right)}-\left[\ve{p}^{\left(\nu\right)}\right]_k \ve{\varphi}_k \right)^\trans \cdot \ve{W} \cdot \ldots \\&& \hspace{2.75cm} \cdot \left( \ve{r}^{\left(\nu-1\right)}-\left[\ve{p}^{\left(\nu\right)}\right]_k \ve{\varphi}_k \right) \bigg) \\
 	\hspace{-1mm}&=&\hspace{-1mm} \argmax_k \frac{\left(\left(\ve{r}^{\left(\nu-1\right)}\right)^\trans \ve{W} \ve{\varphi}_k\right)^2}{\ve{\varphi}_k^\trans \ve{W} \ve{\varphi}_k} \\
 	\hspace{-1mm}&=&\hspace{-1mm} \argmax_k\left[ \ve{p}^{\left(\nu\right)} \diag\left(\ve{p}^{\left(\nu\right)}\right) \diag\left(\ve{\Phi}^\trans \ve{W} \ve{\Phi}\right) \right]_k.
\end{eqnarray}
}{
\begin{eqnarray}
 	u \hspace{-1mm}&=&\hspace{-1mm} \argmin_k   \bigg(\left( \ve{r}^{\left(\nu-1\right)}-\left[\ve{p}^{\left(\nu\right)}\right]_k \ve{\varphi}_k \right)^\trans \cdot \ve{W} \cdot \left( \ve{r}^{\left(\nu-1\right)}-\left[\ve{p}^{\left(\nu\right)}\right]_k \ve{\varphi}_k \right) \bigg) \\
 	\hspace{-1mm}&=&\hspace{-1mm} \argmax_k \frac{\left(\left(\ve{r}^{\left(\nu-1\right)}\right)^\trans \ve{W} \ve{\varphi}_k\right)^2}{\ve{\varphi}_k^\trans \ve{W} \ve{\varphi}_k} \\
 	\hspace{-1mm}&=&\hspace{-1mm} \argmax_k\left[ \ve{p}^{\left(\nu\right)} \diag\left(\ve{p}^{\left(\nu\right)}\right) \diag\left(\ve{\Phi}^\trans \ve{W} \ve{\Phi}\right) \right]_k.
\end{eqnarray}
}
Here, the fact is exploited that the vector $\left(\ve{r}^{\left(\nu-1\right)}-\left[\ve{p}^{\left(\nu\right)}\right]_k \ve{\varphi}_k\right)$ is orthogonal to $\ve{\varphi}_k$ with respect to the weighted scalar product
\begin{equation}
	\left(\ve{r}^{\left(\nu-1\right)}-\left[\ve{p}^{\left(\nu\right)}\right]_k \ve{\varphi}_k\right)^\trans \ve{W} \ve{\varphi}_k \exeq 0.
\end{equation}

\begin{figure}
 	\centering
\psfrag{ph1}[c][c][0.9]{$\ve{\varphi}_1$}
\psfrag{ph2}[c][c][0.9]{$\ve{\varphi}_2$}
\psfrag{ph3}[c][c][0.9]{$\ve{\varphi}_3$}
\psfrag{c1}[c][c][0.9]{\color{red}$c_1$}
\psfrag{c2}[c][c][0.9]{\color{red}$c_2$}
\psfrag{c3}[c][c][0.9]{\color{red}$c_3$}
\psfrag{p1}[c][c][0.9]{\color{blue}$p_1$}
\psfrag{r}[l][l][0.9]{$\ve{r}=c_1\ve{\varphi}_1 + c_2\ve{\varphi}_2 + c_3\ve{\varphi}_3$}
\ifthenelse{\equal{\isfinal}{true}}
{\includegraphics[width=0.25\textwidth]{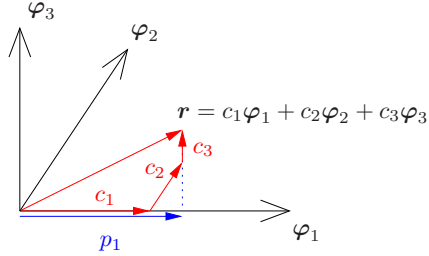}}
{\includegraphics[width=0.35\textwidth]{graphics/od_example.eps}}
\caption{Example for orthogonality deficiency: projection coefficient $p_1$ is larger than original coefficient $c_1$ since portion of $\ve{\varphi}_2$ in direction of $\ve{\varphi}_1$ also contributes to the projection.}
\label{fig:ortho_deficiency}
\end{figure}

In the next step, the coefficient vector $\hat{\ve{c}}$ has to be updated. The update should cover the original portion the selected basis vector $\ve{\varphi}_u$ has of the residual vector $\ve{r}^{\left(\nu-1\right)}$. Unfortunately, due to the loss, the original portion could not be recovered exactly. However, the weighted projection of $\ve{r}^{\left(\nu-1\right)}$ onto the basis vector $\ve{\varphi}_u$ leads to a good initial estimate for the original portion and could be used for updating the coefficient vector. The original 3D-FSE algorithm from \cite{Meisinger2007} directly uses the result from the weighted projection as estimate for the weight of the corresponding basis vector. But by using the result from the weighted projection directly, the update suffers from orthogonality deficiency, as has been shown in detail in \cite{Seiler2007} for the two-dimensional case. The problem is, that although the basis vectors are mutually orthogonal, they are not orthogonal anymore, if the scalar product is evaluated only over support volume $\mathcal{A}$ and in combination with weighting matrix $\ve{W}$. Due to this deficiency of orthogonality, the weighted projection of $\ve{r}^{\left(\nu-1\right)}$ onto the basis vector $\ve{\varphi}_u$ contains portions of basis vectors unlike $\ve{\varphi}_u$ as well. In Fig.\ \ref{fig:ortho_deficiency} this circumstance is presented for a simple example. There, the residual vector $\ve{r}$ emanates from the superposition of $\ve{\varphi}_1$, $\ve{\varphi}_2$ and $\ve{\varphi}_3$ with original weights $c_1$, $c_2$ and $c_3$. In the regarded iteration, basis vector $\ve{\varphi}_1$ is determined to be added to the model. Now, the goal is to derive a good estimate for $c_1$ from the projection coefficients. Apparently, directly using $p_1$ as estimate for $c_1$ is not optimal, since the basis vector $\ve{\varphi}_1$ would become overemphasized in the model. In order to cope with this, in \cite{Seiler2007} we proposed an algorithm using all projection coefficients for compensating the orthogonality deficiency and for obtaining good estimates in the context of two-dimensional extrapolation. Unfortunately, this approach is computationally very expensive. But the analyses presented in \cite{Seiler2008} showed that it is as well sufficient to use a simple orthogonality deficiency compensation which is carried out by only adding a fraction of the projection coefficient to the coefficient vector in a certain iteration. According to \cite{Seiler2008}, the utilization of this simple orthogonality deficiency compensation has two advantages over the original, uncompensated model generation. Firstly, the extrapolation quality can be significantly improved as the risk of overemphasizing a basis vector is effectively eliminated. Secondly, the original Frequency Selective Extrapolation suffers from the problem, that the best possible extrapolation quality is achieved only for a certain number of iterations, which furthermore depends on the content. If the number of performed iterations exceeds this point, the quality degrades very fast. But, if orthogonality deficiency compensation is applied, the extrapolation quality runs into saturation with an increasing number of iterations. Hence, the threat of missing the peak extrapolation quality can be avoided. Utilizing orthogonality deficiency compensation only has one small drawback: the number of iterations which is necessary for model generation has to be increased, compared to the model generation without compensation. Since the advantages of the simple orthogonality deficiency compensation exceed the solely increased number of iterations, this approach is applied to the three-dimensional model generation in the following.

With that, entry $\left[\hat{\ve{c}}^{\left(\nu\right)}\right]_u$ of the coefficient vector in the $\nu$-th iteration corresponding to the selected basis vector $\ve{\varphi}_u$ is updated as follows:
\begin{equation}
 	\left[\hat{\ve{c}}^{\left(\nu\right)}\right]_u = \left[\hat{\ve{c}}^{\left(\nu-1\right)}\right]_u + \gamma \left[\ve{p}^{\left(\nu\right)}\right]_u
\end{equation}
The factor $\gamma$ is from the range between $0$ and $1$ and causes that only the mentioned fraction of the weighted projection's result gets added to the coefficient vector. As has also been shown in \cite{Seiler2008}, fortunately the exact value of $\gamma$ is not very critical. In the case, that the fraction of the basis vector is estimated too small, the same basis vector gets selected again in a later iteration.

After the coefficient vector has been updated, the new residual
\begin{eqnarray}
 	\ve{r}^{\left(\nu\right)} &=& \ve{f} - \ve{\Phi}\hat{\ve{c}}^{\left(\nu\right)} \\
 	 &=& \ve{r}^{\left(\nu-1\right)} - \gamma \left[\ve{p}^{\left(\nu\right)}\right]_u \ve{\varphi}_u
\end{eqnarray}
 is calculated. Then, the new residual $\ve{r}^{\left(\nu\right)}$ is used for further updating the model. Altogether, the above described steps of selecting one basis vector to be added and calculating the corresponding weight are repeated until a predefined number of iterations is reached. 
 
 Finally, $\ve{g}$ is retransformed to the three-dimensional model $g\left[m,n,p\right]$. The values of $g\left[m,n,p\right]$ corresponding to the lost samples from area $\mathcal{B}$ are cut out of the model and are used for replacing the lost signal parts and therewith conceal the error. 
 
In order to indicate the utilization of orthogonality deficiency compensation and to distinguish the novel model generation from the original 3D-FSE from \cite{Meisinger2007}, the enhanced version is referred to as Orthogonality Deficiency Compensated Three-Dimensional Frequency Selective Extrapolation (3D-FSE-OD).
 

\subsection{Motion Compensated Three-Dimensional Frequency Selective Extrapolation Principles} 
\label{ssec:mc_fse}

\begin{figure*}
 	\centering
\ifthenelse{\equal{\isfinal}{true}}
{
	\psfrag{x}[c][c][0.9]{$x$}
	\psfrag{y}[c][c][0.9]{$y$}
	\psfrag{t}[c][c][0.9]{$t$}
	\psfrag{m}[c][c][0.9]{$m$}
	\psfrag{n}[c][c][0.9]{$n$}
	\psfrag{p}[c][c][0.9]{$p$}
	\psfrag{x0}[c][c][0.7]{$x_0$}
	\psfrag{y0}[c][c][0.7]{$y_0$}
	\psfrag{ttau}[c][c][0.7]{$t=\tau$}
	\psfrag{ttaum1}[c][c][0.7]{$t=\tau-1$}
	\psfrag{ttaum2}[c][c][0.7]{$t=\tau-2$}
	\psfrag{ttaum3}[c][c][0.7]{$t=\tau-3$}
	\psfrag{ttaup1}[c][c][0.7]{$t=\tau+1$}
	\psfrag{dtm1}[c][c][0.9]{\color{red} $\ve{d}^{\left(-1\right)}$}
	\psfrag{dtm2}[c][c][0.9]{\color{red} $\ve{d}^{\left(-2\right)}$}
	\psfrag{dtm3}[c][c][0.9]{\color{red} $\ve{d}^{\left(-3\right)}$}
	\psfrag{dtp1}[c][c][0.9]{\color{red} $\ve{d}^{\left(+1\right)}$}
	\psfrag{Loss area}[l][l][0.9]{Loss area $\mathcal{B}$}
	\psfrag{Support volume}[l][l][0.9]{Support volume $\mathcal{A}$}
	\psfrag{Video sequence}[l][l][0.9]{Video sequence}
	\psfrag{Motion compensated and}[l][l][0.9]{Motion compensated and}
	\psfrag{aligned extrapolation volume}[l][l][0.9]{aligned extrapolation volume}
 	\includegraphics[width=0.83\textwidth]{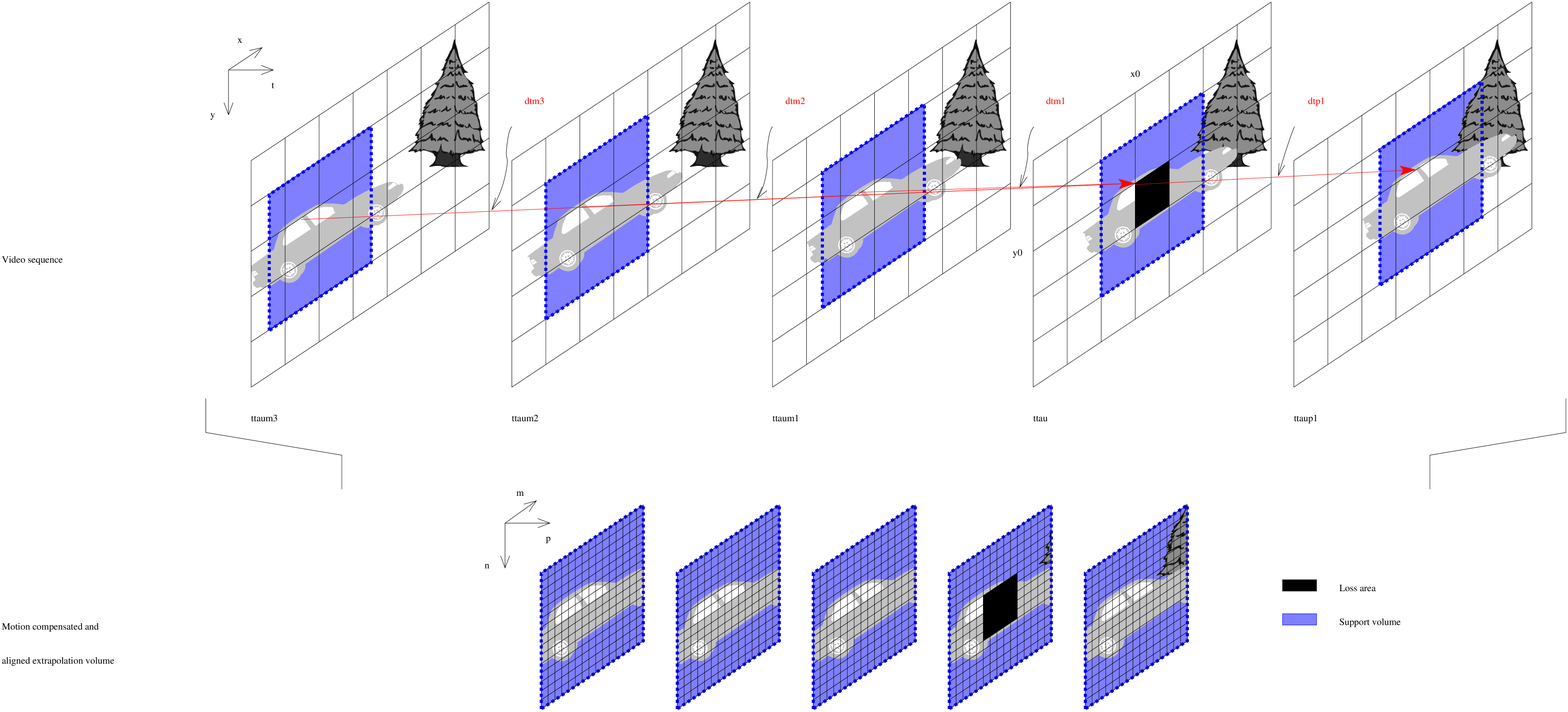}
}{ 	
	\psfrag{x}[c][c][0.7]{$x$}
	\psfrag{y}[c][c][0.7]{$y$}
	\psfrag{t}[c][c][0.7]{$t$}
	\psfrag{m}[c][c][0.7]{$m$}
	\psfrag{n}[c][c][0.7]{$n$}
	\psfrag{p}[c][c][0.7]{$p$}
	\psfrag{x0}[c][c][0.5]{$x_0$}
	\psfrag{y0}[c][c][0.5]{$y_0$}
	\psfrag{ttau}[c][c][0.5]{$t=\tau$}
	\psfrag{ttaum1}[c][c][0.5]{$t=\tau-1$}
	\psfrag{ttaum2}[c][c][0.5]{$t=\tau-2$}
	\psfrag{ttaum3}[c][c][0.5]{$t=\tau-3$}
	\psfrag{ttaup1}[c][c][0.5]{$t=\tau+1$}
	\psfrag{dtm1}[c][c][0.7]{\color{red} $\ve{d}^{\left(-1\right)}$}
	\psfrag{dtm2}[c][c][0.7]{\color{red} $\ve{d}^{\left(-2\right)}$}
	\psfrag{dtm3}[c][c][0.7]{\color{red} $\ve{d}^{\left(-3\right)}$}
	\psfrag{dtp1}[c][c][0.7]{\color{red} $\ve{d}^{\left(+1\right)}$}
	\psfrag{Loss area}[l][l][0.7]{Loss area $\mathcal{B}$}
	\psfrag{Support volume}[l][l][0.7]{Support volume $\mathcal{A}$}
	\psfrag{Video sequence}[l][l][0.7]{Video sequence}
	\psfrag{Motion compensated and}[l][l][0.7]{Motion compensated and}
	\psfrag{aligned extrapolation volume}[l][l][0.7]{aligned extrapolation volume}
 	\includegraphics[width=0.95\textwidth]{graphics/mcfse-extrapolation_area_2.eps}
} 	
 	\caption{Top: distorted video sequence $v\left[x,y,t\right]$ with isolated block loss in frame $t=\tau$ at $\left(x_0,y_0\right)$. Bottom: aligned extrapolation volume $\mathcal{L}_\mathrm{a}$.} 	
 	\label{fig:mcfse_extrapolation_area}
\end{figure*}

Regarding Fig.\ \ref{fig:3dfse_extrapolation_area} closely, it becomes obvious, that the distorted object is moving and is at different positions in the individual frames. The movement from frame $t=\tau$ to frame $t=\tau+\kappa$ can be described by the motion vector
\begin{equation}
 	\ve{d}^{\left(\kappa\right)} = \left(x_\mathrm{d}^{\left(\kappa\right)}, y_\mathrm{d}^{\left(\kappa\right)}\right)
\end{equation}
where $x_\mathrm{d}^{\left(\kappa\right)}$ is the displacement in horizontal and $y_\mathrm{d}^{\left(\kappa\right)}$ in vertical direction. As can be seen in Fig.\ \ref{fig:3dfse_extrapolation_area} bottom, due to the motion, the fixed data volume used for the original \mbox{3D-FSE} and the enhanced 3D-FSE-OD cannot be optimal for the model generation, since the signal parts that are lost in the current frame are at different positions in the previous or future frames. Even worse, it may be possible that they are not part of a fixed volume at all. 

To overcome this weakness, the proposed MC-FSE aims at revoking the motion of the object in order to get an aligned volume before the model generation takes place. In doing so, Fig.\ \ref{fig:mcfse_extrapolation_area} top shows the same video sequence as before, but there the extrapolation volume is aligned according to the motion. In \mbox{Fig.\ \ref{fig:mcfse_extrapolation_area}} bottom, the aligned extrapolation volume $\mathcal{L}_\mathrm{a}$ is presented for this example. Obviously, such an aligned extrapolation volume is much more suited for the subsequent model generation as the different layers of the volume, corresponding to the considered previous and future frames, contain the same signal parts as the layer corresponding to the frame suffering from the loss. In addition to that, every layer contains the region corresponding to the loss at the same position and therewith the lost area can be reconstructed more precisely. The aligned extrapolation volume is also depicted by coordinates $m$, $n$ and $p$ and is of size $M\times N\times P$.  Volume $\mathcal{L}_\mathrm{a}$ again can be split into two parts, region $\mathcal{B}$ subsuming the lost samples and support volume $\mathcal{A}$, subsuming the correctly received samples. 

\begin{figure}
 	\centering
 	\ifthenelse{\equal{\isfinal}{true}}
	{\psfrag{Lost block}[c][c][0.8]{Lost block}
 	\psfrag{Fractional-pel}[c][c][0.8]{Fractional-pel}
	\psfrag{motion estimation}[c][c][0.8]{motion estimation}
	\psfrag{Reliability check}[c][c][0.8]{Reliability check}
	\psfrag{Estimation}[c][c][0.8]{Estimation}
	\psfrag{reliable?}[c][c][0.8]{reliable?}
	\psfrag{yes}[c][c][0.8]{yes}
	\psfrag{no}[c][c][0.8]{no}
	\psfrag{Volume alignment}[c][c][0.8]{Volume alignment}
	\psfrag{Model generation}[c][c][0.8]{3D Model generation}
	\psfrag{Cut block out of the model}[c][c][0.8]{Cut block out of the model}
	\psfrag{Replace lost block}[c][c][0.8]{Replace lost block}
	\includegraphics[width=0.23\textwidth]{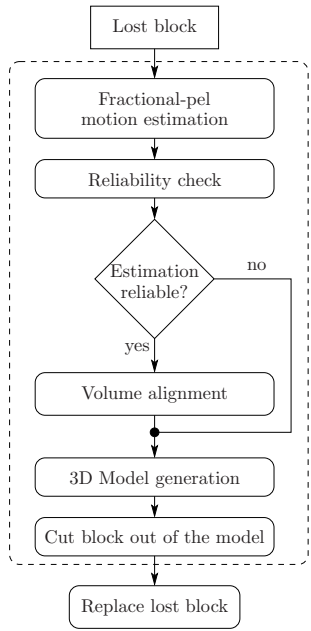}}
	{\psfrag{Lost block}[c][c][0.7]{Lost block}
 	\psfrag{Fractional-pel}[c][c][0.7]{Fractional-pel}
	\psfrag{motion estimation}[c][c][0.7]{motion estimation}
	\psfrag{Reliability check}[c][c][0.7]{Reliability check}
	\psfrag{Estimation}[c][c][0.7]{Estimation}
	\psfrag{reliable?}[c][c][0.7]{reliable?}
	\psfrag{yes}[c][c][0.7]{yes}
	\psfrag{no}[c][c][0.7]{no}
	\psfrag{Volume alignment}[c][c][0.7]{Volume alignment}
	\psfrag{Model generation}[c][c][0.7]{3D Model generation}
	\psfrag{Cut block out of the model}[c][c][0.7]{Cut block out of the model}
	\psfrag{Replace lost block}[c][c][0.7]{Replace lost block}
	\includegraphics[width=0.33\textwidth]{graphics/block_diagram_MC-FSE.eps}}
	\caption{Block diagram for Motion Compensated Three-Dimensional Frequency Selective Extrapolation.}
	\label{fig:block_diagramm_mcfse}
\end{figure}

Fig.\ \ref{fig:block_diagramm_mcfse} shows a block diagram of the different steps of \mbox{MC-FSE} that are required for obtaining the aligned volume and for extrapolating the signal. Unlike \cite{Seiler2008a}, where only full-pel accuracy is used for motion estimation, we propose to estimate the motion in the sequence around the lost block with fractional-pel accuracy in the first step. Since motion estimation may be inaccurate, its reliability has to be checked. If the estimation is considered to be reliable, the extrapolation volume is aligned according to the estimated fractional-pel positions before the above described 3D-FSE-OD is applied on the aligned volume for generating the model. Otherwise, the estimation is discarded and \mbox{3D-FSE-OD} is applied on the fixed volume, centered by the lost block. After the model has been generated, again the samples corresponding to the lost block are cut out of the model and are used for concealing the loss.

\subsection{Motion Estimation}
\label{ssec:MC-FSE_ME}

As has been shown in \cite{Ericsson1985, Girod1987, Girod1993}, video sequences can be compressed more efficiently by using motion compensated prediction with fractional-pel accuracy. This is due to the fact that the motion in a sequence does not necessarily have to follow the artificially introduced sampling grid of the sensor. Although fractional-pel motion estimation is quite ordinary in encoding video sequences it is used only very rarely for error concealment. The algorithms proposed in \cite{Fumagalli2006} and \cite{Park1994} are two of the few examples for using fractional-pel motion estimation, but the vast majority of extrapolation algorithms only uses full-pel accuracy. Since error concealment can be regarded as a signal extrapolation task closely related to motion compensated prediction, the idea of fractional-pel motion estimation is used subsequently. For this, the undistorted previous and future frames have to be upsampled. To simplify matters, only integer scaling factors are considered. The upsampling factor is denoted by $D$. The upsampling is performed by first inserting $D-1$ zeros between every two neighboring samples. Afterwards, these zeros are removed by interpolating the signal between the original samples. The most simple interpolation would be just using bilinear filtering. Alternatively, more sophisticated interpolation schemes like the two-stage interpolation used in H.264/AVC \cite{ISO/IEC2003} or the adaptive interpolation proposed in \cite{Wedi2006} can be used. Subsuming all regarded frames, we get the upsampled sequence $v_\mathrm{u}\left[x,y,t\right]$ of spatial size $\left(D\left(X-1\right)+1\right) \times \left(D\left(Y-1\right)+1\right)$. Thereby, the original full-pel samples get mapped from spatial coordinates $\left(x,y\right)$ to $\left(Dx, Dy\right)$.

\begin{figure}
 	\centering
	\psfrag{x0}[c][c][0.9]{$x_0$}
	\psfrag{y0}[c][c][0.9]{$y_0$}
	\psfrag{M}[c][c][0.9]{$\mathcal{M}$}
	\ifthenelse{\equal{\isfinal}{true}}
	{\includegraphics[width=0.4\textwidth]{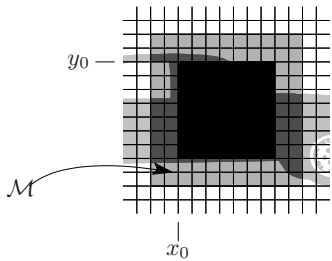}}
	{\includegraphics[width=0.5\textwidth]{graphics/motion_estimation.eps}}
	\caption{Decision area $\mathcal{M}$ around the lost block at $\left(x_0,y_0\right)$.}
	\label{fig:motion_estimation}
\end{figure}

After the upsampling of the reference frames, the motion in the sequence around the lost block is estimated. Here, the limitation has to be considered, that the samples of the lost block cannot be taken into account for estimating the motion. To cope with this, the same idea is used as is utilized in DMVE \cite{Zhang2000a} for recovering the lost motion vector. There, an area around the lost block is regarded and motion estimation is performed for this area, meaning that the best matching area in the regarded frame is determined. Fig.\ \ref{fig:motion_estimation} shows the relation between the so called decision area $\mathcal{M}$ and the lost block. To estimate the motion of the object between the frame at $t=\tau$ and the upsampled frame at $t=\tau+\kappa$, all motion vectors up to a maximum displacement of $d_\mathrm{max}$ are tested. For this, for every candidate motion vector $\widetilde{\ve{d}}^{\left(\kappa\right)} = \left(\widetilde{x}_\mathrm{d}^{\left(\kappa\right)}, \widetilde{y}_\mathrm{d}^{\left(\kappa\right)}\right)$ the sum of squared errors
\ifthenelse{\equal{\isfinal}{true}}
{\[
 E^{\left(\kappa \right)}\left(\tilde{\ve{d}}^{\left(\kappa \right)}\right) = \hspace{6.5cm}
\]
\begin{equation}
\sum_{\forall \left(x,y\right) \in \mathcal{M}} \hspace{-0.4cm} \left(v\left[x,y,\tau\right] \hspace{-0.1cm}-\hspace{-0.1cm} v_\mathrm{u}\left[Dx+\tilde{x}^{\left(\kappa \right)}_\mathrm{d}, Dy+\tilde{y}^{\left(\kappa \right)}_\mathrm{d}, \tau+\kappa\right] \right)^2
\label{eq:dmve_error}
\end{equation}}
{\begin{equation}
E^{\left(\kappa \right)}\left(\tilde{\ve{d}}^{\left(\kappa \right)}\right) = \sum_{\forall \left(x,y\right) \in \mathcal{M}} \hspace{-0.4cm} \left(v\left[x,y,\tau\right] \hspace{-0.1cm}-\hspace{-0.1cm} v_\mathrm{u}\left[Dx+\tilde{x}^{\left(\kappa \right)}_\mathrm{d}, Dy+\tilde{y}^{\left(\kappa \right)}_\mathrm{d}, \tau+\kappa\right] \right)^2
\label{eq:dmve_error}
\end{equation}}
between all samples from area $\mathcal{M}$ in frame $t=\tau$ and the displaced version of $\mathcal{M}$ in frame $t=\tau+\kappa$ is calculated. If a candidate vector points outside the margins of the examined frame, the regarded frame is padded respectively with the samples of its margin. $\kappa$ indicates all the frames used for concealment and is from the range $-N_\mathrm{p}, \ldots, N_\mathrm{f} \backslash 0$.

After the sum of squared errors has been determined for every candidate motion vector, the one is selected, that minimizes the error, resulting in
\begin{eqnarray}
	\hat{\ve{d}}^{\left(\kappa\right)} & = & \hspace{-0.75cm} \argmin_{\parbox{2.5cm}{\centering$\widetilde{\ve{d}}^{\left(\kappa\right)}$}} \hspace{-0.25cm} E^{\left(\kappa \right)}\left(\tilde{\ve{d}}^{\left(\kappa \right)}\right)	\\ 
	& = & \hspace{-0.75cm} \argmin_{\parbox{2.5cm}{\centering \tiny$\tilde{x}^{\left(\kappa \right)}_\mathrm{d}, \tilde{y}^{\left(\kappa \right)}_\mathrm{d} = -d_\mathrm{max}, \ldots, d_\mathrm{max}$}} \hspace{-0.25cm} E^{\left(\kappa \right)}\left(\left(\tilde{x}^{\left(\kappa \right)}_\mathrm{d}, \tilde{y}^{\left(\kappa \right)}_\mathrm{d}\right)\right).
\end{eqnarray}
Since the reliability of the estimation is checked in the next step, the estimation error 
\begin{equation}
\check{E}^{\left(\kappa \right)} = E^{\left(\kappa \right)}\left( \hat{\ve{d}}^{\left(\kappa \right)}\right)
\end{equation}
belonging to the selected motion vector is stored for further usage. 

These steps are performed for all considered previous and future frames. This results in a set of estimated motion vectors $\hat{\ve{d}}^{\left(\kappa\right)}$ with $\kappa=-N_\mathrm{p}, \ldots, N_\mathrm{f} \backslash 0$ and the estimation errors $\check{E}^{\left(\kappa \right)}$ corresponding to each of the vectors. 

\subsection{Motion Estimation Evaluation}

The above mentioned reliability check for the motion estimation quality is important since motion cannot be estimated well in some cases, e.\ g., if non-translational movements or content changes occur. In such a case, the estimated motion vectors do not fit the real motion in the scene and an extrapolation volume $\mathcal{L}_\mathrm{a}$ aligned according to those vectors is less suited for the subsequent model generation than an unaligned one. In order to evaluate the estimation reliability, two thresholds are used for checking the absolute quality of the estimation and its homogeneity. In the case that at least one of the two thresholds is exceeded, the estimation is discarded and the model generation is performed on the unaligned, fixed volume.

For testing the absolute motion estimation quality, the maximum of the square root of the estimation errors $\check{E}^{\left(\kappa \right)}$, normalized by the number of samples in $\mathcal{M}$ is used. Thus, the maximum mean error per pixel is calculated for the motion vectors to all regarded frames. If this value is above threshold $T_\mathrm{abs}$, \mbox{i.\ e.\ }
\begin{equation}
 	\max_{\kappa=-N_\mathrm{p},\ldots,N_\mathrm{f}\backslash 0} \hspace{-1mm} \sqrt{\frac{\check{E}^{\left(\kappa \right)}}{\left|\mathcal{M}\right|}}> T_\mathrm{abs},
\end{equation}
the motion estimation is regarded as unreliable. For checking the homogeneity of the estimation, the difference between the maximum and the minimum estimation error is calculated and afterwards normalized by the mean estimation error of all frames. This quotient
\begin{equation}
 	\frac{\displaystyle \max_{\kappa=-N_\mathrm{p},\ldots,N_\mathrm{f}\backslash 0} \hspace{-1mm} \sqrt{\check{E}^{\left(\kappa \right)}} - \min_{\kappa=-N_\mathrm{p},\ldots,N_\mathrm{f}\backslash 0} \hspace{-1mm} \sqrt{\check{E}^{\left(\kappa \right)}} }{ \displaystyle \frac{1}{N_\mathrm{p}+N_\mathrm{f}} \sum_{\kappa=-N_\mathrm{p},\ldots,N_\mathrm{f}\backslash 0} \sqrt{\check{E}^{\left(\kappa \right)}} }> T_\mathrm{rel}
\end{equation}
is compared to the second threshold $T_\mathrm{rel}$. Again, if this threshold is exceeded, the motion estimation for this volume is decided to be unreliable and is discarded.

\subsection{Volume Alignment}

After the motion in the sequence has been estimated and the estimation has been evaluated, the extrapolation volume $\mathcal{L}_\mathrm{a}$ and therewith $f\left[m,n,p\right]$ is set up. In the case that the estimation is regarded as unreliable, the extrapolation volume consists of the samples from frame $t=\tau$ belonging to the lost block, a border of $B$ samples width surrounding the loss, and the samples at the same spatial positions in the considered previous and future frames. Thus, signal $f\left[m,n,p\right]$ in this volume emanates from the video sequence according to:
\begin{equation}
	f\left[m,n,p\right] = v\left[x_0-B+m, y_0-B+n, \tau-N_\mathrm{p}+p\right].
\end{equation}

If the estimation is classified as reliable, from the frame at $t=\tau$ the same samples as above are taken for the extrapolation volume. But, in order to take previous and future frames into account, the samples are taken from the upsampled sequence $v_\mathrm{u}\left[x,y,t\right]$, shifted according to the estimated motion. In doing so, the areas from the previous and future frames can be positioned with fractional-pel accuracy and therewith the real motion of an object can be taken into account more precisely. With that, signal $f\left[m,n,p\right]$ in extrapolation volume $\mathcal{L}_\mathrm{a}$ is given by:
\begin{equation}
 f\left[m,n,p\right] = v\left[x_0-B+m, y_0-B+n, \tau\right] \mbox{ , for } p=N_\mathrm{p}
\end{equation}
and for $ p=0, \ldots, \left(N_\mathrm{p}+N_\mathrm{f}\right) \backslash N_\mathrm{p}$:
\ifthenelse{\equal{\isfinal}{true}}{
\begin{eqnarray}
 \nonumber f\left[m,n,p\right] &=& v_\mathrm{u}\Big[D\left(x_0-B\right) + \hat{x}_\mathrm{d}^{\left(p-N_\mathrm{p}\right)} +Dm , \\\nonumber &&\hspace{5mm} D\left(y_0-B\right) + \hat{y}_\mathrm{d}^{\left(p-N_\mathrm{p}\right)} +Dn , \\&&\hspace{5mm} \tau-N_\mathrm{p}+p\Big]
\end{eqnarray}}
{\[
  f\left[m,n,p\right]= \hspace{10cm}
 \]
 \begin{equation}
  v_\mathrm{u}\Big[D\left(x_0-B\right) + \hat{x}_\mathrm{d}^{\left(p-N_\mathrm{p}\right)} +Dm ,  D\left(y_0-B\right) + \hat{y}_\mathrm{d}^{\left(p-N_\mathrm{p}\right)} +Dn , \tau-N_\mathrm{p}+p\Big].
 \end{equation}
}

In the next step, the model of the signal in the aligned volume $\mathcal{L}_\mathrm{a}$ is generated by means of 3D-FSE-OD in order to extrapolate the signal from the correctly received areas into the lost area. Since the motion is revoked by aligning the extrapolation volume prior to the model generation the shortcoming of 3D-FSE, respectively 3D-FSE-OD is resolved and an improved extrapolation quality can be achieved. Finally, the samples of the model corresponding to loss area $\mathcal{B}$ are cut out of the model and are used for concealing the distortion.


\section{Results}
\label{sec:results}

\subsection{Simulation Setup}
For testing the abilities of MC-FSE, we implemented the algorithm into the H.264/AVC decoder of reference software JM14.0 \cite{JVT2008}. An overview of the H.264/AVC codec can be found in \cite{Wiegand2003,Sullivan2005,Ostermann2004}. In addition to MC-FSE , Temporal Replacement (TR) \cite{Bopardikar2005}, EBMA \cite{Lam1993}, DMVE \cite{Zhang2000a}, MH-EC \cite{Song2007}, IFS \cite{Hwang2008}, the original 3D-FSE \cite{Meisinger2007} and the enhanced 3D-FSE-OD are implemented in order to compare the performance with other error concealment algorithms. None of the regarded algorithms makes use of an explicit scene change detection. In general, all temporal and spatio-temporal error concealment algorithms suffer from scene changes, as in such a case the reference frames can contain different content compared to the distorted frame. But as we focus on examining the extrapolation quality in our experiments and want to avoid influences from a scene change detection algorithm and intra-frame concealment, which becomes necessary if scene changes occur, we do not include a scene change detection in our experiments. In addition to that, the examined sequences only contain few scene changes and this handicap is present for all the regarded algorithms in the same way.

\begin{figure}
 	\centering
 	\psfrag{Slice group 1}[l][l][0.7]{Slice group 1}
 	\psfrag{Slice group 2}[l][l][0.7]{Slice group 2}
 	\psfrag{Slice group 3}[l][l][0.7]{Slice group 3}
 	\psfrag{Slice group 4}[l][l][0.7]{Slice group 4}
 	\psfrag{b}[c][c][1.]{b)}
 	\ifthenelse{\equal{\isfinal}{true}}
	{\includegraphics[width=0.48\textwidth]{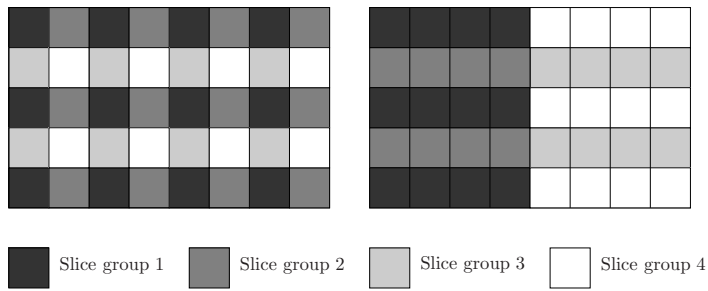}}
	{\includegraphics[width=0.75\textwidth]{graphics/fmo_patterns.eps}}
 	\caption{H.264/AVC FMO patterns. Left: Dispersed. Right: Interleaved.}
 	\label{fig:fmo_patterns}
\end{figure}

\begin{figure}
 	\centering
 	\psfrag{a}[c][c][1.]{a)}
 	\psfrag{b}[c][c][1.]{b)}
 	\ifthenelse{\equal{\isfinal}{true}}
	{\includegraphics[width=0.48\textwidth]{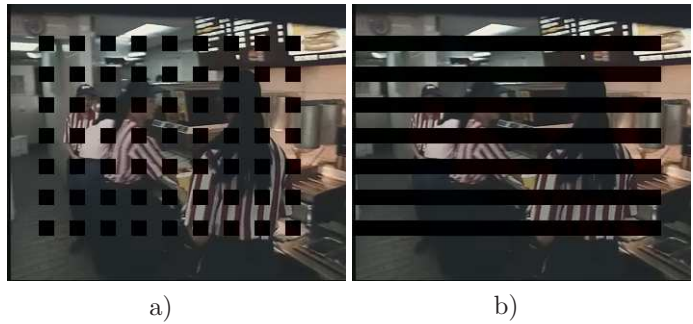}}
	{\includegraphics[width=0.75\textwidth]{graphics/error_pattern.eps}}
 	\caption{Examined error patterns: a) DISPERSED, b) INTERLEAVED.}
 	\label{fig:error_pattern}
\end{figure}

For simulating the losses and obtaining the results listed in the subsequent tables, three different error patterns are imprinted on the just decoded frames at the decoder. These losses are concealed by the various error concealment algorithms. If no error concealment is applied, the distorted and uncorrected frame is copied to the buffer of decoded frames, causing severe error propagation. Fig.\ \ref{fig:visual_results} b) shows an example image, suffering from the imprinted error pattern as well as error propagation from previous frames. We prefer this artificial method over using a channel simulator on the coded bitstream since inserting errors into a coded bitstream produces unpredictable errors in the decoded frames. Changing a bit in the data stream of one sequence might cause that a whole frame cannot be decoded at all whereas in another sequence a flipped bit at the same position only corrupts one single block. Since we aim at evaluating the actual signal extrapolation performance within the scope of our research, having the same error patterns for the different examined sequences is indispensable. Nevertheless, we choose the error patterns to be close to typical errors occurring if an H.264/AVC coded bitstream gets distorted. As shown in \cite{Stockhammer2005}, H.264/AVC offers the possibility to divide a frame into several slices that can be encoded and decoded independently. The standard further contains Flexible Macroblock Ordering (FMO) which controls the assignment of macroblocks to the different slices. Fig.\ \ref{fig:fmo_patterns} shows two typical FMO patterns consisting of four slice groups. If the bits of one of the slice groups gets distorted by a transmission error, the macroblocks contained in this slice group become corrupted. Considering the FMO patterns from Fig.\ \ref{fig:fmo_patterns}, a transmission error results either in isolated block losses or consecutive block losses. Hence, the first error pattern that we consider is called DISPERSED (see Fig.\ \ref{fig:error_pattern} a) ) and consits of isolated block losses of size $16\times 16$ samples, occurring with a distance of $16$ samples between them. The second error pattern is called INTERLEAVED (see \mbox{Fig.\ \ref{fig:error_pattern} b) )} and is used for simulating consecutive block losses where several lines of macroblocks get lost within one frame, either located at the left side of the frame or shifted to the right. The third error pattern is called MIXED and consits of isolated as well as consecutive block losses. In every other frame, the error patterns are shifted in vertical and horizontal direction. For the evaluation, the CIF-sequences ``Vimto'', ``Hockey'', ``Basketball'', ``Canoe'', ``Fast Food'' and ``Foreman'' are regarded. The sequences are coded with $\punit{QP} 28$ in IPPP order with a group of pictures (GOP) length of $12$ frames for concealing errors in P-frames and in IPBP order with a GOP length of $23$ frames for concealing errors in B-frames. The sequences have a length of $216$ frames or $184$ frames, respectively. Thereby, the errors are imprinted either in the P-frames or in the B-frames only. Regarding errors in B-frames, the special case arises, that no error propagation can occur and past and future frames can be used for concealing the losses. For evaluating the quality of the concealment, the Peak Signal to Noise Ratio ($\PSNR$) of the luminance component is used as measure. In this context, $\PSNR$ is evaluated against the coded sequence and only over regions that are suffering either from the imprinted errors or from error propagation. In doing so, the extrapolation quality can be evaluated more accurately since the undistorted samples are not taken into account. Although $\PSNR$ is not a good metric to evaluate the absolute quality of a sequence, it is suited well for comparing the relative quality. According to \cite{Oelbaum2007}, $\PSNR$ is highly correlated to visual quality if similar distortions are regarded within one sequence. So, if for one sequence an algorithm can achieve a higher $\PSNR$ compared to another one, this also corresponds to an improved visual quality of the concealment.  

The motion estimation for MC-FSE is performed with full-pel, half-pel and quarter-pel accuracy. If half-pel and quarter-pel accuracy is used, the twice and fourfold upsampled reference frames are generated by the same procedure as used in H.264/AVC. There, the half-pel positions are calculated by using a \mbox{$6$-tap} filter in horizontal and vertical direction. The quarter-pel positions are calculated by bilinear interpolation between the two nearest full- and half-pel samples. DMVE and EBMA use a full search for the candidate blocks and the search is carried out also with full-pel, half-pel, and quarter-pel accuracy in the same way as for MC-FSE in order to achieve a fair comparison. 

\subsection{Parameter Selection}
For carrying out the simulations the following parameter set is used. Most of the parameters that are necessary for the model generation for 3D-FSE, 3D-FSE-OD and MC-FSE are selected according to the originally proposed 3D-FSE in \cite{Meisinger2007}. Hence, the used basis functions are the functions of the three-dimensional discrete Fourier transform. As has been shown in detail in \cite{Kaup2005} and \cite{Meisinger2007}, this set of basis functions is suited well for extrapolating all kinds of content, smooth as well as noise like areas and edges. In addition to that, an efficient implementation in the transform domain can be used for these functions. For this, an FFT of size $64\times 64\times 16$ is used for the transform into the Fourier domain. The support volume $\mathcal{A}$ has a width of $B=16$ samples, and as far as available $N_\mathrm{p}=2$ previous frames are used for reference when concealing \mbox{P-frames} and, if possible, $N_\mathrm{p}=2$ previous and $N_\mathrm{f}=1$ following frames are used for reference when concealing \mbox{B-frames}. The decay factor of the weighting function is set to $\hat{\rho} = 0.8$, and if consecutive block losses are concealed, the already concealed blocks are weighted by the additional factor $\delta = 0.2$. Whereas only $200$ iterations are carried out for 3D-FSE for generating the model, the number of iterations for 3D-FSE-OD and MC-FSE is set to $800$. The increased number of iterations is necessary as 3D-FSE-OD and MC-FSE make use of orthogonality deficiency compensation which requires more iterations for generating the model. In this context, the orthogonality deficiency compensation factor $\gamma$ is set to $0.7$. Selecting these values for $\gamma$ and the number of iterations, a good trade-off between extrapolation quality and computational complexity can be met. As shown in \cite{Seiler2008}, the extrapolation quality could be further improved slightly by using smaller values for $\gamma$ and increasing the number of iterations, but we abstained from this option in order to keep the computational load manageable.

For all algorithms that make use of motion estimation, a full search is performed with a maximum search range of $d_\mathrm{max}=16$ samples. The decision area $\mathcal{M}$ used for \mbox{MC-FSE} is a border of $4$ samples width. As shown in \cite{Zhang2000a}, a small width results in a low complexity but also a reduced estimation accuracy. In contrast to this, a large width increases the estimation accuracy but also increases the computational load. Thus, we use a border width of $4$ samples as a trade-off between estimation quality and computational complexity. 

The two thresholds $T_\mathrm{rel}$ and $T_\mathrm{abs}$ which MC-FSE uses for evaluating the motion estimation are set to $T_\mathrm{rel} = 3$ and $T_\mathrm{abs}=10$. For deriving these parameters, we have tested a lot of different parameter sets on the sequences ``Discovery City'' and ``Discovery Orient''. These sequences are suited well, as they contain very different scenes, some with fast and some with slow motion, as well as some with homogenous and others with textured content. Fortunately, the exact values of the listed parameters are not very critical. That is to say, the parameters can be varied widely without affecting the extrapolation quality heavily and none of the parameters depends on certain sequences. To give an example, if the parameters $T_\mathrm{rel}$ and $T_\mathrm{abs}$ are varied between $1.5<T_\mathrm{rel}<4.5$ and $5<T_\mathrm{abs}<20$ respectively, the extrapolation quality for error pattern DISPERSED in P-frames for these two sequences only varies between $-0.06\punit{dB}$ and $+0.19\punit{dB}$. But if parameters far away from the above listed ones are regarded, the extrapolation quality drops significantly. If very large values are considered, the motion vectors are selected every time and a misalignment of the extrapolation cannot be prevented, resulting in a reduced extrapolation quality. But if very small values are considered, the motion estimation is always regarded as unreliable and the performance of MC-FSE drops to the level of 3D-FSE-OD. Hence, selecting the parameters from the above given ranges provides a good trade-off between the risk of misaligning the extrapolation volume and effectively revoking the estimated motion for setting up a more suited extrapolation volume.

\begin{table*}
\renewcommand{\arraystretch}{1.075}
\ifthenelse{\equal{\isfinal}{true}}{
\small
\caption{Extrapolation quality for $216$ Frames, GOP size $12$, P-Frames, $\punit{QP} 28$ and error pattern DISPERSED}
\begin{center}
\begin{tabular}{|p{2.25cm}|C{1.75cm}|C{1.75cm}|C{1.75cm}|C{1.75cm}|C{1.75cm}|C{1.75cm}|}}
{\tiny
\caption{Extrapolation quality for $216$ Frames, GOP size $12$, P-Frames, $\punit{QP} 28$ and error pattern DISPERSED}
\begin{center}
\begin{tabular}{|l|c|c|c|c|c|c|}}
\hline & ``Vimto'' & ``Hockey'' & ``Basketball'' & ``Canoe'' & ``Fast Food'' & ``Foreman''\\ \hline\hline
TR \cite{Bopardikar2005} & $21.59 \punit{dB}$ & $17.70 \punit{dB}$ & $15.17 \punit{dB}$ & $19.07 \punit{dB}$ & $17.35 \punit{dB}$ & $24.07 \punit{dB}$\\ \hline
DMVE \cite{Zhang2000a} & $25.78 \punit{dB}$ & $23.51 \punit{dB}$ & $20.07 \punit{dB}$ & $22.69 \punit{dB}$ & $23.02 \punit{dB}$ & $31.85 \punit{dB}$\\ \hline
DMVE HPel & $25.71 \punit{dB}$ & $23.77 \punit{dB}$ & $20.49 \punit{dB}$ & $22.90 \punit{dB}$ & $23.23 \punit{dB}$ & $32.58 \punit{dB}$\\ \hline
DMVE QPel & $25.86 \punit{dB}$ & $24.00 \punit{dB}$ & $20.81 \punit{dB}$ & $23.27 \punit{dB}$ & $23.39 \punit{dB}$ & $33.08 \punit{dB}$\\ \hline
EBMA \cite{Lam1993} & $27.38 \punit{dB}$ & $23.30 \punit{dB}$ & $19.11 \punit{dB}$ & $22.81 \punit{dB}$ & $23.43 \punit{dB}$ & $28.46 \punit{dB}$\\ \hline
EBMA HPel & $27.49 \punit{dB}$ & $23.47 \punit{dB}$ & $19.37 \punit{dB}$ & $22.99 \punit{dB}$ & $23.57 \punit{dB}$ & $28.42 \punit{dB}$\\ \hline
EBMA QPel & $27.54 \punit{dB}$ & $23.73 \punit{dB}$ & $19.82 \punit{dB}$ & $23.34 \punit{dB}$ & $23.83 \punit{dB}$ & $28.72 \punit{dB}$\\ \hline
MH-EC \cite{Song2007} & $27.09 \punit{dB}$ & $24.33 \punit{dB}$ & $20.89 \punit{dB}$ & $23.59 \punit{dB}$ & $24.06 \punit{dB}$ & $33.07 \punit{dB}$\\ \hline
IFS \cite{Hwang2008} & $29.28 \punit{dB}$ & $24.41 \punit{dB}$ & $20.92 \punit{dB}$ & $24.31 \punit{dB}$ & $24.66 \punit{dB}$ & $30.45 \punit{dB}$\\ \hline
3D-FSE \cite{Meisinger2007} & $29.15 \punit{dB}$ & $26.52 \punit{dB}$ & $20.85 \punit{dB}$ & $23.74 \punit{dB}$ & $25.05 \punit{dB}$ & $32.88 \punit{dB}$\\ \hline
3D-FSE-OD & $29.99 \punit{dB}$ & $27.30 \punit{dB}$ & $21.66 \punit{dB}$ & $24.71 \punit{dB}$ & $25.93 \punit{dB}$ & $33.63 \punit{dB}$\\ \hline
MC-FSE & $30.42 \punit{dB}$ & $27.22 \punit{dB}$ & $21.69 \punit{dB}$ & $24.70 \punit{dB}$ & $25.96 \punit{dB}$ & $33.29 \punit{dB}$\\ \hline
MC-FSE HPel & $30.49 \punit{dB}$ & $27.57 \punit{dB}$ & $21.76 \punit{dB}$ & $24.78 \punit{dB}$ & $26.07 \punit{dB}$ & $34.26 \punit{dB}$\\ \hline
MC-FSE QPel & $30.58 \punit{dB}$ & $27.69 \punit{dB}$ & $21.89 \punit{dB}$ & $24.82  \punit{dB} $ & $26.09 \punit{dB}$ & $34.69 \punit{dB}$\\ \hline
\end{tabular}
\label{tab:p-dispersed_all}
\end{center}
\end{table*}

\begin{table*}
\renewcommand{\arraystretch}{1.075}
\ifthenelse{\equal{\isfinal}{true}}{
\small
\caption{Mean gain vs. DMVE for different error scenarios}
\begin{center}
\begin{tabular}{|p{2.cm}|C{2.cm}|C{2.cm}|C{2.cm}|C{2.cm}|C{2.cm}|C{2.cm}|}
 \hline & \multicolumn{3}{c|}{P-Concealment} & \multicolumn{3}{c|}{B-Concealment} \\
& \multicolumn{1}{c}{DISPERSED} & \multicolumn{1}{c}{INTERLEAVED} & \multicolumn{1}{c|}{MIXED} & \multicolumn{1}{c}{DISPERSED} & \multicolumn{1}{c}{INTERLEAVED} & \multicolumn{1}{c|}{MIXED}\\ \hline \hline
}
{\tiny
\caption{Mean gain vs. DMVE for different error scenarios}
\begin{center}
\begin{tabular}{|p{1.5cm}|c|c|c|c|c|c|}}
\hline & \multicolumn{3}{c|}{P-Concealment} & \multicolumn{3}{c|}{B-Concealment} \\
& \multicolumn{1}{c}{\hspace{-1mm}DISPERSED} & \multicolumn{1}{c}{\hspace{-2mm}INTERLEAVED} & \multicolumn{1}{c|}{\hspace{-3mm}MIXED} & \multicolumn{1}{c}{\hspace{-1mm}DISPERSED} & \multicolumn{1}{c}{\hspace{-2mm}INTERLEAVED} & \multicolumn{1}{c|}{\hspace{-3mm}MIXED}\\ \hline \hline
TR \cite{Bopardikar2005}	& $-5.33 \punit{dB}$	& $-2.85 \punit{dB}$	& $-4.38 \punit{dB}$	& $-5.34 \punit{dB}$	& $-3.05 \punit{dB}$ & $-4.79 \punit{dB}$\\ \hline
DMVE \cite{Zhang2000a}		& ---	& ---	& ---	& ---	& --- & --- \\ \hline
DMVE HPel			& $0.29 \punit{dB}$	& $0.18 \punit{dB}$	& $0.26 \punit{dB}$	& $0.33 \punit{dB}$	& $0.23 \punit{dB}$ & $0.34 \punit{dB}$\\ \hline
DMVE QPel			& $0.58 \punit{dB}$	& $0.44 \punit{dB}$	& $0.57 \punit{dB}$	& $0.53 \punit{dB}$	& $0.38 \punit{dB}$ & $0.54 \punit{dB}$\\ \hline
EBMA \cite{Lam1993}		& $-0.40 \punit{dB}$	& $-0.74 \punit{dB}$	& $-1.31 \punit{dB}$	& $-0.95 \punit{dB}$	& $-1.15 \punit{dB}$ & $-2.15\punit{dB}$\\ \hline
EBMA HPel			& $-0.27 \punit{dB}$	& $-0.69 \punit{dB}$	& $-1.26 \punit{dB}$	& $-0.74 \punit{dB}$	& $-1.03 \punit{dB}$ & $-1.97 \punit{dB}$\\ \hline
EBMA QPel			& $0.01 \punit{dB}$	& $-0.52 \punit{dB}$	& $-1.04 \punit{dB}$	& $-0.55 \punit{dB}$	& $-0.84 \punit{dB}$ & $-1.79 \punit{dB}$\\ \hline
MH-EC \cite{Song2007}		& $1.02 \punit{dB}$	& $0.82 \punit{dB}$	& $0.44 \punit{dB}$	& $0.70 \punit{dB}$	& $0.68 \punit{dB}$ & $0.06 \punit{dB}$\\ \hline
IFS \cite{Hwang2008}		& $1.19 \punit{dB}$	& $1.27 \punit{dB}$	& $0.64 \punit{dB}$	& $1.04 \punit{dB}$	& $0.89 \punit{dB}$ & $-0.03 \punit{dB}$\\ \hline
3D-FSE \cite{Meisinger2007}	& $1.88 \punit{dB}$	& $2.01  \punit{dB}$	& $1.14 \punit{dB}$	& $2.87 \punit{dB}$	& $2.94 \punit{dB}$ & $1.76 \punit{dB}$\\ \hline
3D-FSE-OD			& $2.72 \punit{dB}$	& $2.77  \punit{dB}$	& $1.97 \punit{dB}$	& $3.77 \punit{dB}$	& $3.88 \punit{dB}$ & $2.66 \punit{dB}$\\ \hline
MC-FSE				& $2.73 \punit{dB}$	& $2.68 \punit{dB}$	& $1.98 \punit{dB}$	& $3.89 \punit{dB}$	& $3.85 \punit{dB}$ & $2.82 \punit{dB}$\\ \hline
MC-FSE HPel			& $3.00 \punit{dB}$	& $3.00 \punit{dB}$	& $2.25 \punit{dB}$	& $4.28 \punit{dB}$	& $4.38 \punit{dB}$ & $3.26 \punit{dB}$\\ \hline
MC-FSE QPel			& $3.14 \punit{dB}$	& $3.16  \punit{dB}$	& $2.39 \punit{dB}$	& $4.42 \punit{dB}$	& $4.49 \punit{dB}$ & $3.40 \punit{dB}$\\ \hline
\end{tabular}
\label{tab:mean_gain}
\end{center}
\end{table*}

\ifthenelse{\equal{\isfinal}{true}}{
\begin{table*}
\renewcommand{\arraystretch}{1.075}
\small
\caption{Mean gain produced by fractional-pel motion estimation}
\begin{center}
\begin{tabular}{|p{4.25cm}|p{3cm}|C{1.75cm}|C{1.75cm}|C{1.75cm}|}
\hline \multicolumn{2}{|c|}{}& DMVE & EBMA & MC-FSE \\ \hline \hline
P-Concealment& Full-Pel 		& $24.49 \punit{dB}$ & $24.08 \punit{dB}$ & $27.21 \punit{dB}$\\ \cline{2-5}
DISPERSED& Gain Half-Pel 		& $0.29 \punit{dB}$ & $0.13 \punit{dB}$ & $0.27 \punit{dB}$ \\ \cline{2-5}
& Gain Quarter-Pel 			& $0.58 \punit{dB}$ & $0.41 \punit{dB}$ & $0.41 \punit{dB}$ \\ \hline\hline
P-Concealment& Full-Pel 		& $20.55 \punit{dB}$ & $19.81 \punit{dB}$ & $23.23 \punit{dB}$ \\\cline{2-5}
INTERLEAVED& Gain Half-Pel 		& $0.18 \punit{dB}$ & $0.05 \punit{dB}$ & $0.32 \punit{dB}$ \\ \cline{2-5}
& Gain Quarter-Pel 			& $0.44 \punit{dB}$ & $0.22 \punit{dB}$ & $0.48 \punit{dB}$ \\ \hline\hline
P-Concealment& Full-Pel 		& $22.58 \punit{dB}$ & $21.27 \punit{dB}$ & $24.55 \punit{dB}$ \\\cline{2-5}
MIXED& Gain Half-Pel 			& $0.26 \punit{dB}$ & $0.05 \punit{dB}$ & $0.27 \punit{dB}$ \\ \cline{2-5}
& Gain Quarter-Pel 			& $0.57 \punit{dB}$ & $0.27 \punit{dB}$ & $0.41 \punit{dB}$ \\ \hline\hline
B-Concealment& Full-Pel 		& $24.91 \punit{dB}$ & $23.96 \punit{dB}$ & $28.81 \punit{dB}$ \\\cline{2-5}
DISPERSED& Gain Half-Pel 		& $0.33 \punit{dB}$ & $0.21 \punit{dB}$ & $0.39 \punit{dB}$ \\ \cline{2-5}
& Gain Quarter-Pel 			& $0.53 \punit{dB}$ & $0.40 \punit{dB}$ & $0.53 \punit{dB}$ \\ \hline\hline
B-Concealment& Full-Pel 		& $22.73 \punit{dB}$ & $21.58 \punit{dB}$ & $26.59 \punit{dB}$  \\\cline{2-5}
INTERLEAVED& Gain Half-Pel 		& $0.23 \punit{dB}$ & $0.12 \punit{dB}$ & $0.53 \punit{dB}$ \\ \cline{2-5}
& Gain Quarter-Pel 			& $0.38 \punit{dB}$ & $0.31 \punit{dB}$ & $0.64 \punit{dB}$ \\ \hline\hline
B-Concealment& Full-Pel 		& $24.39 \punit{dB}$ & $22.25 \punit{dB}$ & $27.22 \punit{dB}$  \\\cline{2-5}
MIXED& Gain Half-Pel 			& $0.34 \punit{dB}$ & $0.18 \punit{dB}$ & $0.44 \punit{dB}$ \\ \cline{2-5}
& Gain Quarter-Pel 			& $0.54 \punit{dB}$ & $0.36 \punit{dB}$ & $0.58 \punit{dB}$ \\ \hline
\end{tabular}
\label{tab:fractional_pel_gain}
\end{center}
\end{table*}}
{
\begin{table*}
\renewcommand{\arraystretch}{1.075}
\tiny
\caption{Mean gain produced by fractional-pel motion estimation}
\begin{center}
\begin{tabular}{|l|l|C{1.25cm}|C{1.25cm}|C{1.25cm}|}
\hline \multicolumn{2}{|c|}{}& DMVE & EBMA & MC-FSE \\ \hline \hline
P-Concealment& Full-Pel 		& $24.49 \punit{dB}$ & $24.08 \punit{dB}$ & $27.21 \punit{dB}$\\ \cline{2-5}
DISPERSED& Gain Half-Pel 		& $0.29 \punit{dB}$ & $0.13 \punit{dB}$ & $0.27 \punit{dB}$ \\ \cline{2-5}
& Gain Quarter-Pel 			& $0.58 \punit{dB}$ & $0.41 \punit{dB}$ & $0.41 \punit{dB}$ \\ \hline\hline
P-Concealment& Full-Pel 		& $20.55 \punit{dB}$ & $19.81 \punit{dB}$ & $23.23 \punit{dB}$ \\\cline{2-5}
INTERLEAVED& Gain Half-Pel 		& $0.18 \punit{dB}$ & $0.05 \punit{dB}$ & $0.32 \punit{dB}$ \\ \cline{2-5}
& Gain Quarter-Pel 			& $0.44 \punit{dB}$ & $0.22 \punit{dB}$ & $0.48 \punit{dB}$ \\ \hline\hline
P-Concealment& Full-Pel 		& $22.58 \punit{dB}$ & $21.27 \punit{dB}$ & $24.55 \punit{dB}$ \\\cline{2-5}
MIXED& Gain Half-Pel 			& $0.26 \punit{dB}$ & $0.05 \punit{dB}$ & $0.27 \punit{dB}$ \\ \cline{2-5}
& Gain Quarter-Pel 			& $0.57 \punit{dB}$ & $0.27 \punit{dB}$ & $0.41 \punit{dB}$ \\ \hline\hline
B-Concealment& Full-Pel 		& $24.91 \punit{dB}$ & $23.96 \punit{dB}$ & $28.81 \punit{dB}$ \\\cline{2-5}
DISPERSED& Gain Half-Pel 		& $0.33 \punit{dB}$ & $0.21 \punit{dB}$ & $0.39 \punit{dB}$ \\ \cline{2-5}
& Gain Quarter-Pel 			& $0.53 \punit{dB}$ & $0.40 \punit{dB}$ & $0.53 \punit{dB}$ \\ \hline\hline
B-Concealment& Full-Pel 		& $22.73 \punit{dB}$ & $21.58 \punit{dB}$ & $26.59 \punit{dB}$  \\\cline{2-5}
INTERLEAVED& Gain Half-Pel 		& $0.23 \punit{dB}$ & $0.12 \punit{dB}$ & $0.53 \punit{dB}$ \\ \cline{2-5}
& Gain Quarter-Pel 			& $0.38 \punit{dB}$ & $0.31 \punit{dB}$ & $0.64 \punit{dB}$ \\ \hline\hline
B-Concealment& Full-Pel 		& $24.39 \punit{dB}$ & $22.25 \punit{dB}$ & $27.22 \punit{dB}$  \\\cline{2-5}
MIXED& Gain Half-Pel 			& $0.34 \punit{dB}$ & $0.18 \punit{dB}$ & $0.44 \punit{dB}$ \\ \cline{2-5}
& Gain Quarter-Pel 			& $0.54 \punit{dB}$ & $0.36 \punit{dB}$ & $0.58 \punit{dB}$ \\ \hline
\end{tabular}
\label{tab:fractional_pel_gain}
\end{center}
\end{table*}}

\subsection{Objective Quality Evaluation}
In Table \ref{tab:p-dispersed_all} the obtainable $\PSNR$ values are listed for concealing error pattern DISPERSED in P-frames by using the above listed error concealment techniques. In order to concentrate the results and to show the other scenarios as well, Table \ref{tab:mean_gain} lists the gain in $\PSNR$ each error concealment technique can achieve against DMVE, averaged over all sequences. Thereby, the error patterns DISPERSED, INTERLEAVED and MIXED are applied on P- and B-frames. Regarding the two tables, it becomes obvious that TR is inferior to all compared algorithms, but it is also by far the most simple one. The two temporal algorithms EBMA and DMVE provide a decent extrapolation quality, while DMVE is superior to EBMA in most cases. This is especially true when regarding error patterns INTERLEAVED and MIXED. In this case, EBMA cannot estimate the motion well and falls behind DMVE. But, both algorithms can also significantly benefit from motion compensation with fractional-pel accuracy. In Table \ref{tab:fractional_pel_gain}, the gain achievable by fractional-pel motion estimation is listed for the different scenarios. So, for DMVE a gain of up to $0.58 \punit{dB}$ can be achieved if motion estimation is performed with quarter-pel accuracy, for EBMA a gain of up to $0.41 \punit{dB}$ is possible.
 
Regarding the spatio-temporal algorithms MH-EC, IFS and 3D-FSE, which are used for reference, one can see that taking the spatial neighborhood into account can significantly improve the extrapolation quality. In this context, 3D-FSE exploits the additional information more effectively than the two other algorithms and outperforms them by $0.50$ to $2.26 \punit{dB}$ $\PSNR$ in the mean. Taking a look at 3D-FSE-OD, it becomes also obvious that the incorporation of orthogonality deficiency compensation into 3D-FSE leads to a further improvement and 3D-FSE-OD is able to outperform the original 3D-FSE by $0.76$ to $0.94 \punit{dB}$. 

When evaluating the gain that can be achieved by estimating the motion and aligning the extrapolation volume, one has to compare 3D-FSE-OD and MC-FSE. In doing so, one can recognize that for P-frames approximately the same performance can be obtained if motion estimation and volume alignment is performed only with full-pel accuracy. This is due to the fact that although the extrapolation quality is improved for most blocks by aligning the volume prior to the model generation, in some cases the volume can get slightly misaligned, despite of the motion estimation evaluation. In such a case, the extrapolation quality of MC-FSE can be inferior to \mbox{3D-FSE-OD} and the improvement gained so far is consumed.
But, by using quarter-pel accuracy the extrapolation quality can be improved in the mean by at least $0.39 \punit{dB}$. Hence, \mbox{MC-FSE} is superior to all other considered error concealment algorithms if half-pel or \mbox{quarter-pel} accuracy is used for motion estimation and volume alignment. The improvement further increases if B-frames are considered. For B-frames, MC-FSE outperforms 3D-FSE-OD by up to $0.74 \punit{dB}$ if quarter-pel accuracy is applied. The gain over DMVE grows to up to $4.49 \punit{dB}$. The very high extrapolation quality of 3D-FSE, 3D-FSE-OD and MC-FSE for B-frames can be explained by the fact that there previous and future frames are available. Due to this, the lost blocks are surrounded by correctly received blocks in spatial and temporal direction which is very helpful for the model generation. Comparing MC-FSE to the original 3D-FSE, a gain of up to $1.64 \punit{dB}$ can be noticed, which arises in B-frames for error pattern MIXED. But, similar gains can be achieved for the other error patterns as well. If the results of 3D-FSE-OD are also taken into account, one can recognize that approximately half the gain is produced by applying orthogonality deficiency compensation to the three-dimensional model generation and the other half is produced by motion estimation and volume alignment with fractional-pel accuracy. In addition to the above mentioned DMVE and EBMA, Table \ref{tab:fractional_pel_gain} lists the gain in extrapolation quality that can be achieved for MC-FSE by using motion estimation with fractional-pel accuracy. Apparently, the gain achievable by fractional-pel motion estimation is similar for MC-FSE as for the aforementioned algorithms and by using half-pel motion estimation, the extrapolation quality can be improved by up to $0.53 \punit{dB}$, by using quarter-pel accuracy by up to $0.64 \punit{dB}$.

\begin{figure}
 	\centering
	\psfrag{s05}[t][t]{\color[rgb]{0,0,0}\setlength{\tabcolsep}{0pt}\begin{tabular}{c}Frame number\end{tabular}}%
	\psfrag{s06}[b][b]{\color[rgb]{0,0,0}\setlength{\tabcolsep}{0pt}\begin{tabular}{c}Discarded motion vectors\end{tabular}}%
	\psfrag{s08}[b][b]{\color[rgb]{0,0,0}\setlength{\tabcolsep}{0pt}\begin{tabular}{c}\end{tabular}}%
	\psfrag{s10}[][]{\color[rgb]{0,0,0}\setlength{\tabcolsep}{0pt}\begin{tabular}{c} \end{tabular}}%
	\psfrag{s11}[][]{\color[rgb]{0,0,0}\setlength{\tabcolsep}{0pt}\begin{tabular}{c} \end{tabular}}%
	\psfrag{s12}[l][l][0.7]{\color[rgb]{0,0,0}}%
	\psfrag{s17}[l][l][0.7]{\color[rgb]{0,0,0}MC-FSE}%
	\psfrag{s18}[l][l][0.7]{\color[rgb]{0,0,0}MC-FSE HPel}%
	\psfrag{s19}[l][l][0.7]{\color[rgb]{0,0,0}MC-FSE QPel}%
	\psfrag{v12}[r][r][1.]{$0$}%
	\psfrag{v13}[r][r][1.]{}
	\psfrag{v14}[r][r][1.]{$20\%$}%
	\psfrag{v15}[r][r][1.]{}
	\psfrag{v16}[r][r][1.]{$40\%$}%
	\psfrag{v17}[r][r][1.]{}
	\psfrag{v18}[r][r][1.]{$60\%$}%
	\psfrag{v19}[r][r][1.]{}
	\psfrag{v20}[r][r][1.]{$80\%$}%
	\psfrag{v21}[r][r][1.]{}
	\psfrag{v22}[r][r][1.]{$100\%$}%
	\psfrag{x12}[t][t]{$20$}%
	\psfrag{x13}[t][t]{$40$}%
	\psfrag{x14}[t][t]{$60$}%
	\psfrag{x15}[t][t]{$80$}%
	\psfrag{x16}[t][t]{$100$}%
	\psfrag{x17}[t][t]{$120$}%
	\psfrag{x18}[t][t]{$140$}%
	\psfrag{x19}[t][t]{$160$}%
	\psfrag{x20}[t][t]{$180$}%
	\psfrag{x21}[t][t]{$200$}%

	\ifthenelse{\equal{\isfinal}{true}}{\includegraphics[width=0.45\textwidth]{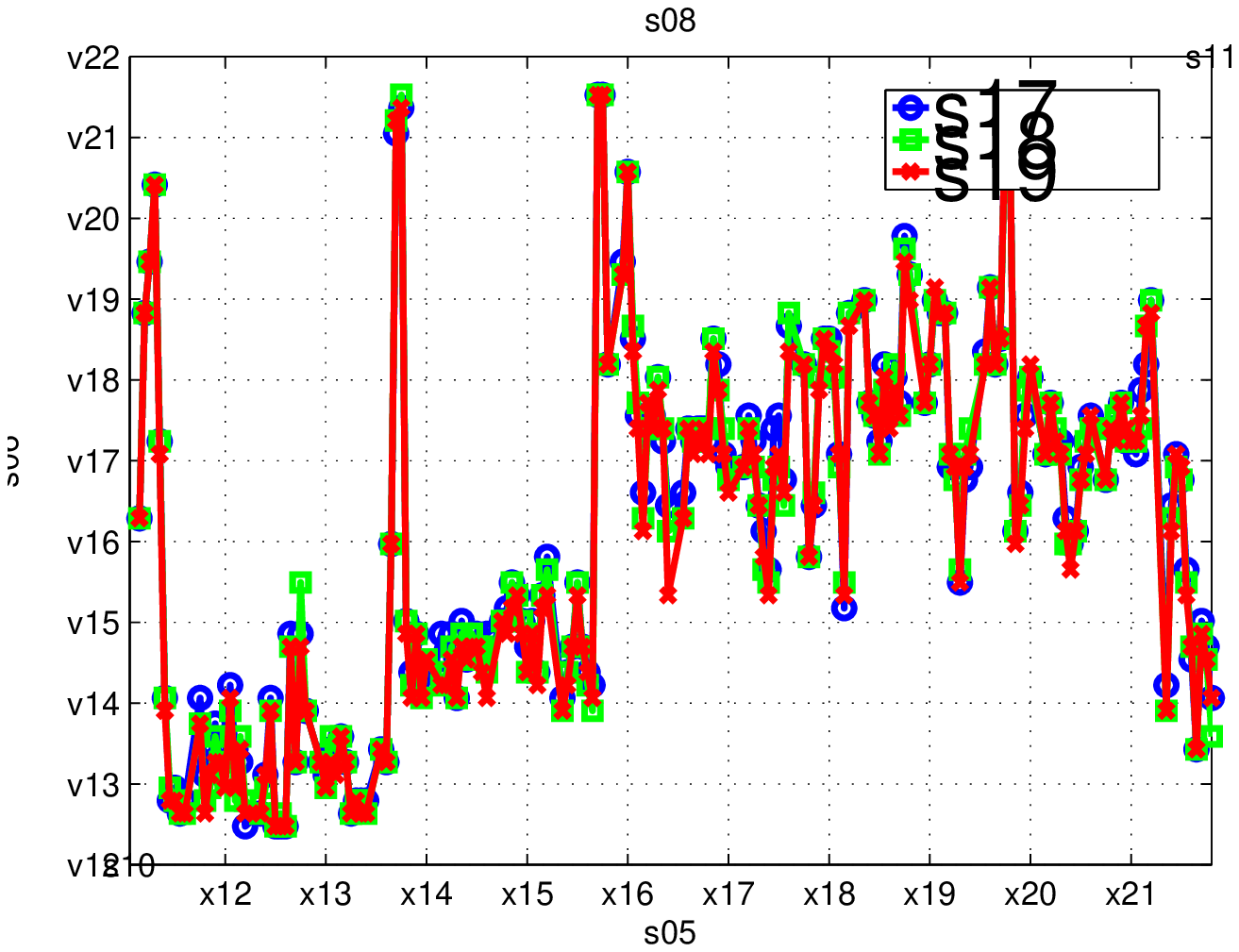}}
	{\includegraphics[width=0.75\textwidth]{graphics/vimto_rejected_mv.eps}}
 	\caption{Percentage of discarded motion vectors per frame for sequence ``Vimto'' distorted by error pattern DISPERSED.}
 	\label{fig:mv_rejection_plot}
\end{figure}

Fig.\ \ref{fig:mv_rejection_plot} shows the percentage of discarded motion vectors per frame for sequence ``Vimto'' with error pattern DISPERSED in P-frames if MC-FSE is carried out with full-pel, half-pel, and quarter-pel accuracy. Apparently, in some frames approximately all the motion vectors are used for aligning the extrapolation volume whereas for other frames nearly all motion vectors are discarded and the fixed extrapolation volume is regarded. The first $93$ frames of sequence ``Vimto'' consist of two scenes with predominant translational motion. During these scenes the motion can be estimated well and be revoked for improving the model generation. At the transition between the two scenes at frame $54$, no reliable motion can be found and MC-FSE falls back to 3D-FSE-OD for almost all blocks, leading to a similar quality for this frame. The second half of the sequence contains a lot of non-translational motion. There, a successful alignment is only possible for few blocks and the motion estimation is dropped in favor of a fixed extrapolation volume. Taking this figure into account it can be discovered that the motion estimation reliability check in MC-FSE behaves as desired and that the alignment of the volume only takes place if it is reasonable.

\begin{table}
\renewcommand{\arraystretch}{1.075}
\ifthenelse{\equal{\isfinal}{true}}{\small}{\scriptsize}
\caption{Extrapolation quality for different $\punit{QP}$s for sequence ``Vimto'' with error pattern DISPERSED in P-Frames}
\begin{center}
\begin{tabular}{|c|c|c|c|c|}
\hline $\punit{QP}$ & 3D-FSE-OD & MC-FSE & MC-FSE & MC-FSE\\&&&HPel & QPel \\ \hline\hline
$13$ & $29.30 \punit{dB}$ & $29.57 \punit{dB}$ & $29.64 \punit{dB}$ & $29.68 \punit{dB}$\\ \hline
$18$ & $29.50 \punit{dB}$ & $29.72 \punit{dB}$ & $29.83 \punit{dB}$ & $29.86 \punit{dB}$\\ \hline
$23$ & $29.72 \punit{dB}$ & $29.95 \punit{dB}$ & $30.09 \punit{dB}$ & $30.12 \punit{dB}$\\ \hline
$28$ & $29.99 \punit{dB}$ & $30.42 \punit{dB}$ & $30.49 \punit{dB}$ & $30.58 \punit{dB}$\\ \hline
$33$ & $30.57 \punit{dB}$ & $30.88 \punit{dB}$ & $30.98 \punit{dB}$ & $30.97 \punit{dB}$\\ \hline
$38$ & $31.30 \punit{dB}$ & $31.69 \punit{dB}$ & $31.73 \punit{dB}$ & $31.75 \punit{dB}$\\ \hline
$43$ & $31.54 \punit{dB}$ & $31.66 \punit{dB}$ & $31.77 \punit{dB}$ & $31.81 \punit{dB}$\\ \hline
\end{tabular}
\label{tab:QP_comparison}
\end{center}
\end{table}

Although the simulation results presented so far are for sequences coded with $\punit{QP} 28$, our experiments have shown that the insights are still valid for stronger and weaker compression levels. To illustrate this, Table \ref{tab:QP_comparison} shows the results for sequence ``Vimto'' coded with different $\punit{QP}$s. Here, again error pattern DISPERSED is concealed in P-frames. Since the $\PSNR$ is evaluated against the undistorted coded sequence, the extrapolation quality slightly increases towards larger $\punit{QP}$s, independently of the concealment algorithm. This is due to the fact that the amount of detail per block decreases due to quantization. The parameters for \mbox{MC-FSE} and 3D-FSE-OD are the same as mentioned above and are not adapted to the individual $\punit{QP}$s as our simulations have shown that an adaption is not required.

\begin{table}
\renewcommand{\arraystretch}{1.075}
\ifthenelse{\equal{\isfinal}{true}}{\small}{\scriptsize}
\caption{Mean extrapolation time per block}
\begin{center}
\begin{tabular}{|l|c|}
\hline Concealment algorithm & Processing time \\ \hline\hline
TR \cite{Bopardikar2005} & $0.004 \punit{sec}$ \\ \hline
DMVE \cite{Zhang2000a} & $0.03 \punit{sec}$ \\ \hline
DMVE HPel & $0.13 \punit{sec}$ \\ \hline
DMVE QPel & $0.50 \punit{sec}$ \\ \hline
EBMA \cite{Lam1993} & $0.02. \punit{sec}$ \\ \hline
EBMA HPel & $0.16 \punit{sec}$ \\ \hline
EBMA QPel & $0.61 \punit{sec}$ \\ \hline
MH-EC \cite{Song2007} & $0.02 \punit{sec}$ \\ \hline
IFS \cite{Hwang2008} & $0.04 \punit{sec}$ \\ \hline
3D-FSE \cite{Meisinger2007} & $1.13 \punit{sec}$ \\ \hline
3D-FSE-OD & $3.72 \punit{sec}$ \\ \hline
MC-FSE & $3.78 \punit{sec}$ \\ \hline
MC-FSE HPel & $3.99 \punit{sec}$ \\ \hline
MC-FSE QPel & $4.91 \punit{sec}$ \\ \hline
\end{tabular}
\label{tab:runtime}
\end{center}
\end{table}

In order to quantify the computational overhead produced by motion estimation and volume alignment and to compare the different algorithms, Table  \ref{tab:runtime} lists the extrapolation time per block for the regarded algorithms. The \mbox{simulations} are carried out on one core of an Intel Core2 Quad @ $2.83\punit{GHz}$. The complete extrapolation process is performed in \mbox{MATLAB} R2007b. To achieve this, the data necessary for concealing a block first is transferred from the JM reference decoder to \mbox{MATLAB} and after extrapolation, the concealed block is retransferred to the reference decoder. Obviously, the algorithms based on Frequency Selective Extrapolation possess an extrapolation time that is significantly higher than the other algorithms. But, as shown above, at the same time, they also provide a significantly improved extrapolation quality. Comparing \mbox{3D-FSE-OD} and MC-FSE with full-pel accuracy, the amount of time to conceal a block is similar. But, if a higher accuracy is used for motion estimation, the number of candidate positions to examine for estimating the motion increases quadratically. For our setup, using quarter-pel accuracy in summary increases the extrapolation time by $32\%$. But, regarding the improved extrapolation quality, this increment should be tolerable.

\subsection{Visual Quality Evaluation}
\begin{figure}
 	\centering
	\psfrag{s01}[t][t]{\color[rgb]{0,0,0}\setlength{\tabcolsep}{0pt}\begin{tabular}{c}Frame number\end{tabular}}%
	\psfrag{s02}[b][b]{\color[rgb]{0,0,0}\setlength{\tabcolsep}{0pt}\begin{tabular}{c}$\PSNR$ in $\punit{dB}$\end{tabular}}%
	\psfrag{s06}[][]{\color[rgb]{0,0,0}\setlength{\tabcolsep}{0pt}\begin{tabular}{c} \end{tabular}}%
	\psfrag{s07}[][]{\color[rgb]{0,0,0}\setlength{\tabcolsep}{0pt}\begin{tabular}{c} \end{tabular}}%
	\psfrag{s13}[l][l][0.7]{\color[rgb]{0,0,0}EBMA QPel}%
	\psfrag{s14}[l][l][0.7]{\color[rgb]{0,0,0}DMVE QPel}%
	\psfrag{s15}[l][l][0.7]{\color[rgb]{0,0,0}MH-EC}%
	\psfrag{s16}[l][l][0.7]{\color[rgb]{0,0,0}IFS}%
	\psfrag{s17}[l][l][0.7]{\color[rgb]{0,0,0}3D-FSE}%
	\psfrag{s18}[l][l][0.7]{\color[rgb]{0,0,0}3D-FSE-OD}%
	\psfrag{s19}[l][l][0.7]{\color[rgb]{0,0,0}MC-FSE QPel}%
	\psfrag{x12}[t][t]{$20$}%
	\psfrag{x13}[t][t]{$40$}%
	\psfrag{x14}[t][t]{$60$}%
	\psfrag{x15}[t][t]{$80$}%
	\psfrag{x16}[t][t]{$100$}%
	\psfrag{x17}[t][t]{$120$}%
	\psfrag{x18}[t][t]{$140$}%
	\psfrag{x19}[t][t]{$160$}%
	\psfrag{x20}[t][t]{$180$}%
	\psfrag{x21}[t][t]{$200$}%
	\psfrag{v12}[r][r]{$20$}%
	\psfrag{v13}[r][r]{$25$}%
	\psfrag{v14}[r][r]{$30$}%
	\psfrag{v15}[r][r]{$35$}%
	\psfrag{v16}[r][r]{$40$}%
	\psfrag{v17}[r][r]{$45$}%

	\ifthenelse{\equal{\isfinal}{true}}{\includegraphics[width=0.45\textwidth]{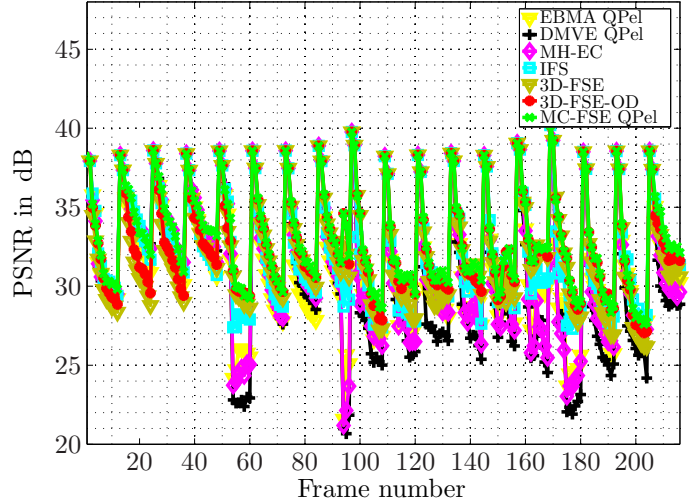}}
	{\includegraphics[width=0.75\textwidth]{graphics/vimto_psnr_vs_frames.eps}}
 	\caption{$\PSNR$ per frame for sequence ``Vimto'' distorted by error pattern DISPERSED. The $\PSNR$ is evaluated for whole frames and against the uncoded original sequence.}
 	\label{fig:psnr_plot}
\end{figure}

In addition to the extrapolation quality of the distorted regions compared to the coded sequence, the quality of the concealed sequence compared to the uncoded sequence is of interest as well, since the consideration of whole frames expresses the viewers impression of a sequence more. To account for this, Fig. \ref{fig:psnr_plot} shows the $\PSNR$ per frame against the uncoded sequence for concealment by EBMA with quarter-pel accuracy, DMVE with quarter-pel accuracy, \mbox{MH-EC}, IFS, 3D-FSE, 3D-FSE-OD, and MC-FSE with quarter-pel accuracy. Here, sequence ``Vimto'' coded with $\punit{QP} 28$ is regarded and error pattern DISPERSED is concealed in P-frames. As there are only errors in P-frames considered, the quality of the concealed sequence is high at the beginning of each GOP and then decreases within the GOP due to error propagation. Furthermore, the scene changes at frame $54$ and $94$ cause a drop of the extrapolation quality, independent of the actual error concealment algorithm. The decay is stronger for temporal error concealment than for spatio-temporal. While spatio-temporal algorithms can also exploit information from the current frame for concealment, temporal ones have to completely rely on previous frames that may be totally distinct in the case of a scene change. Regarding the model generation based algorithms, the extrapolation quality of \mbox{3D-FSE-OD} and MC-FSE is similar for many frames, but there are also a lot of frames where MC-FSE is superior. This happens mainly in frames with high motion. There, the gain of MC-FSE over 3D-FSE and 3D-FSE-OD often grows to more than $2 \punit{dB}$.

\begin{figure*}
 	\centering
 	\includegraphics[width=0.98\textwidth]{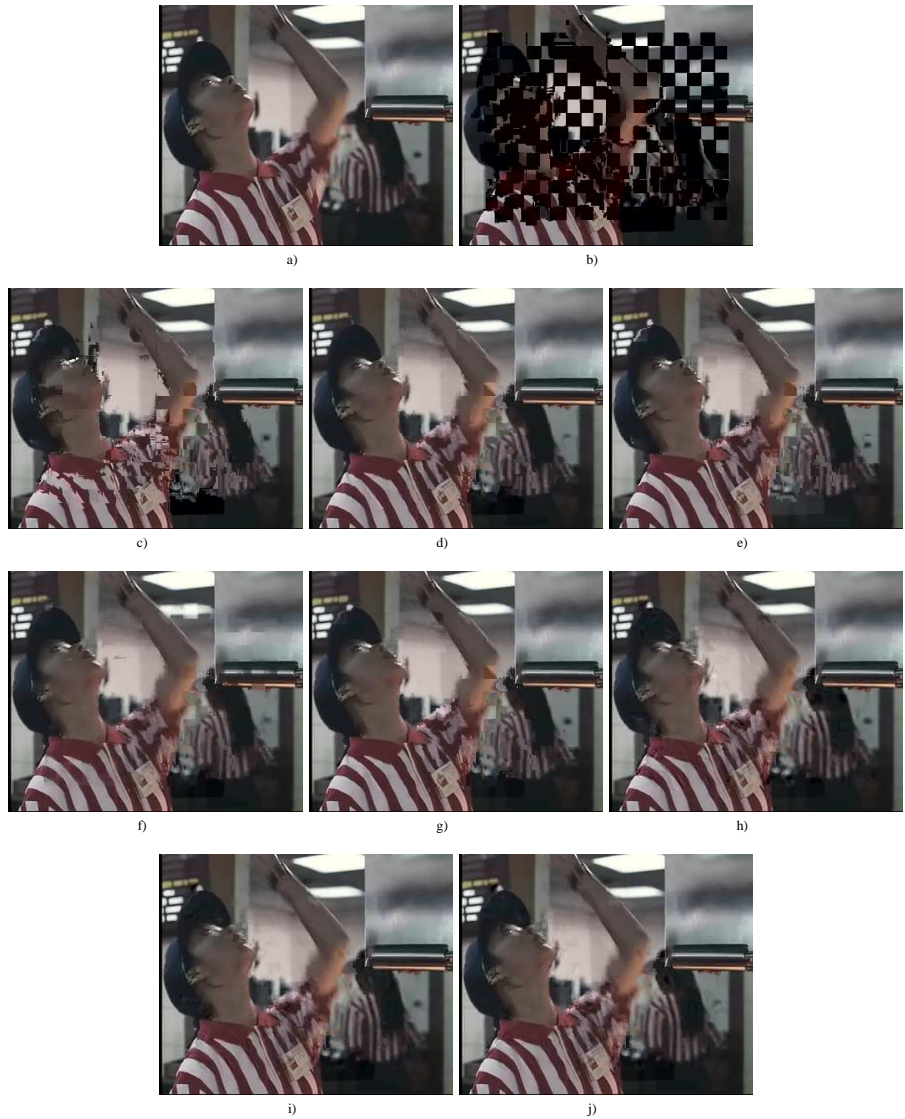}
 	\caption{Visual error concealment results for frame $90$ of test sequence ``Fast Food''. a) Undistorted compressed frame, b) Frame suffering from distortion and error propagation. Error concealment by c) TR, d) EBMA QPel, e) DMVE QPel, f) IFS, g) MH-EC, h) 3D-FSE, i) 3D-FSE-OD, j) MC-FSE QPel.}
 	\label{fig:visual_results}
\end{figure*}

For error concealment, besides the objective measure the visual extrapolation quality is important. In order to illustrate the visual quality, Fig.\ \ref{fig:visual_results} shows frame $90$ of test sequence ``Fast Food'' with different concealment techniques applied to. Subfigure a) shows the undistorted frame, whereas b) shows the frame for which no error concealment is applied. Hence, the frame is suffering from the imprinted error pattern and from error propagation from previous frames. In subfigures c) to e) TR, EBMA and DMVE are used for error concealment. Here, motion compensation with quarter-pel accuracy is used for the latter two. Subfigures f) to h) result from application of IFS, MH-EC and the original 3D-FSE as concealment algorithms. The last two subfigures i) and j) result from error concealment by the proposed 3D-FSE-OD and \mbox{MC-FSE} with quarter-pel motion estimation. Obviously, the frames concealed by TR, EBMA, DMVE are heavily suffering from block artifacts. This is due to the fact that the woman's arm is moving very fast and that EBMA and DMVE cannot find a fitting block in the previous frame. But, the frames concealed by IFS and MH-EC are also suffering from block artifacts and the losses cannot be concealed well. Regarding 3D-FSE, 3D-FSE-OD and \mbox{MC-FSE} which utilize the three-dimensional model generation, no block artifacts can be found. The visual extrapolation quality of these three algorithms is very high and is superior to the other considered algorithms. However, in direct comparison, the frame resulting from MC-FSE shows less artifacts and more details compared to the ones resulting from 3D-FSE-OD. \mbox{3D-FSE-OD} itself provides a superior visual quality compared to 3D-FSE which does not make use of orthogonality deficiency compensation. Altogether, the visual results are reflective for the above outlined $\PSNR$ measurements.

\section{Conclusion}

In this contribution, Motion Compensated Three-Dimensional Frequency Selective Extrapolation was presented for error concealment in video communication. This algorithm utilizes Three-Dimensional Frequency Selective Extrapolation which is further enhanced by orthogonality deficiency compensation for generating a model of the signal suffering from distortion. In doing so, the signal is extrapolated from correctly received regions into the distorted regions. In advance of the model generation, the motion in a video sequence is estimated, evaluated, and compensated with fractional-pel accuracy. The introduced algorithm provides a very high objective as well as subjective extrapolation quality and is able to outperform the original Three-Dimensional Frequency Selective Extrapolation error concealment algorithm significantly. The proposed extension of orthogonality deficiency compensation to the three-dimensional model generation and especially incorporating motion compensation into the model generation yields a gain of up to $1.64 \punit{dB}$ $\PSNR$. Regarding the motion estimation accuracy, a gain of up to $0.64 \punit{dB}$ can be achieved by performing motion estimation with fractional-pel accuracy instead of full-pel accuracy.  

We also showed that typical error concealment algorithms like the Decoder Motion Vector Estimation or the Extended Boundary Matching Algorithm can be improved by up to $0.58 \punit{dB}$ $\PSNR$ by using motion estimation with fractional-pel accuracy if the algorithms are carried out with a full search. 

Besides error concealment in video sequences, the proposed algorithm could be used as well for concealing errors in multi-view sequences in the future. There, in addition to motion, the disparity between different views should compensated as well if the Four-Dimensional Frequency Selective Extrapolation from \cite{Fecker2008} is used. Since disparity as well as the here regarded motion does not necessarily have to follow the sampling grid, the alignment of the extrapolation volume in temporal and view direction can also also be improved by using fractional-pel accuracy. A more precisely aligned four-dimensional hyper-volume can serve as better basis for the model generation and therewith increase the concealment quality.


\end{document}